\documentstyle[12pt]{article}

\def\amsans{y }%
\read-1 to\answ%
\ifx\answ\amsans
\makeatletter
\ifcase\@ptsize
  \font\tenmsy=msbm10
  \font\sevenmsy=msbm7
  \font\fivemsy=msbm5
\or
  \font\tenmsy=msbm10 scaled \magstephalf
  \font\sevenmsy=msbm8
  \font\fivemsy=msbm6
\or
  \font\tenmsy=msbm10 scaled \magstep1
  \font\sevenmsy=msbm8
  \font\fivemsy=msbm6
\fi
\newfam\msyfam
\textfont\msyfam=\tenmsy
\scriptfont\msyfam=\sevenmsy
\scriptscriptfont\msyfam=\fivemsy
\def\Bbb{\ifmmode\let\next\Bbb@\else
\def\next{\errmessage{Use \string\Bbb\space only in math mode}}\fi\next}
\def\Bbb@#1{{\Bbb@@{#1}}}
\def\Bbb@@#1{\fam\msyfam#1}
\newfam\euffam
\font\sixeuf=eufm6
\font\eighteuf=eufm8
\font\twelveeuf=eufm10 scaled\magstep1
\textfont\euffam=\twelveeuf
\scriptfont\euffam=\eighteuf
\scriptscriptfont\euffam=\sixeuf
\def\euf{\fam\euffam\twelveeuf}
\makeatother

\newcommand{\BN}{{\Bbb{N}}}

\newcommand{\BR}{{\Bbb{R}}}
\newcommand{\BZ}{{\Bbb{Z}}}
\newcommand{\Bid}{1\!{\rm l}}
\def\EH{{\euf H}}
\def\EF{{\euf F}}
\newcommand{\myboldmath}{\boldmath}
\else
\def\BR{{\rm I\!R}}
\def\BN{{\rm I\!N}}

\def\BZ{{\rm Z\!\!Z}}

\def\Bid{1\!{\rm l}}
\def\euf{\bf}
\def\EH{{\cal H}}
\def\EF{{\cal F}}
\newcommand{\myboldmath}{\bf}
\fi

\def\pn{\par\noindent}
\def\pano{\par\noindent}

\def\vac#1{|{#1}\rangle}
\def\avac#1{\langle{#1}|}
\def\vak#1{|{\bf#1}\rangle}

\def\vev#1#2{\langle{#1}|{#2}\rangle}

\def\w{{\cal W}}

\def\eps{\varepsilon}

\def\be{\begin{equation}}
\def\ee{\end{equation}}
\def\ba{\begin{array}}
\def\ea{\end{array}}
\def\bea{\begin{eqnarray}}
\def\eea{\end{eqnarray}}
\def\bean{\begin{eqnarray*}}
\def\eean{\end{eqnarray*}}
\def\bl{\begin{list}{}{}}

\def\ds{\displaystyle}
\def\ts{\textstyle}

\newcommand{\reseteqn}{\setcounter{equation}{0}}
\newcommand{\mysection}{\reseteqn\section}
\renewcommand{\thefootnote}{\fnsymbol{footnote}}

\if@twoside
\else
\fi
\begin{document}
  \pagestyle{empty}
  \begin{raggedleft}
CSIC-IMAFF-42-1995\\
arch-ive/9509166\\
September 1995\\
  \end{raggedleft}
%  \begin{center}
%{\eurm DRAFT}
%  \end{center}
  $\phantom{x}$ %%% \vskip 0.618cm\par
  {\LARGE\bf
  \begin{center}
On Modular Invariant Partition Functions\\
of Conformal Field Theories\\
with Logarithmic Operators
  \end{center}
  }\par
  \vfill
  \begin{center}
$\phantom{X}$\\
{\Large Michael A.I.~Flohr\footnote[1]{email: {\tt iffflohr@roca.csic.es}}}\\
{\em Instituto de Matem{\'a}ticas y F{\'\i}sica Fundamental, C.S.I.C.,\\
Serrano 123, E-28006 Madrid, Spain}
  \end{center}\par
  \vfill
  \begin{abstract}
  \noindent
  We extend the definitions of characters and partition functions to the
  case of conformal field theories which contain operators with
  logarithmic correlation functions. As an example we consider the theories
  with central charge $c=c_{p,1}=13-6(p+p^{-1})$, the ``border'' of the
  discrete minimal series. We show that there is a slightly generalized form
  of the property of {\em rationality\/} for such logarithmic theories.
  In particular, we obtain a classification of theories with $c=c_{p,1}$ which
  is similar to the {\em A-D-E\/} classification of $c=1$ models.
  \end{abstract}
  \vfill
  \newpage
%
%%< INTRODUCTION >%%%%%%%%%%%%%%%%%%%%%%%%%%%%%%%%%%%%%%%%%%%%%%%%%%%%%%%%
%
  \setcounter{page}{1}
  \pagestyle{plain}
  \renewcommand{\thefootnote}{\roman{footnote}}
  \mysection{Introduction}
  \pn
This paper mainly deals with conformal field theories (CFT) with central
charge $c=c_{p,1}=1 - 6\frac{(p-1)^2}{p}$, $p\in\BN$, i.e.\ we consider
the ``boundary'' of the minimal discrete series.
  \par
One particularly interesting CFT is the theory with
central charge $c=-2$. It appears in the theoretical treatment of
two-dimensional polymers and self avoiding walks \cite{DuSa87,Sal92},
in the quantum
Hall effect \cite{WeWu94}, and in the phenomenon of unifying $\w$-algebras
\cite{BEHHH94}. It is the first member of a unique series of extended conformal
algebras generated by a triplet of primary fields \cite{Kau91}, which have
central charge $c=c_{p,1}$, $p\in\BN$, and -- last
but not least -- it is the simplest example of a theory containing logarithmic
operators \cite{Gur93}.
  \par
Nonetheless, not very much is known on the representation theory of this
CFT or the other members of the $c_{p,1}$ series. Moreover, while it has been
explicitly shown for $p=2,3$
that there are only finitely many irreducible highest weight representations
with respect to the maximally extended chiral symmetry algebra \cite{EFHHV93,
EHH93b}, the characters and partition functions could not been derived yet.
  \par
The aim of this letter is to extend the definitions of characters and partition
functions such that they cover the case of logarithmic operators, which
necessarily appear in a CFT, whenever the differential equations arising from
the conformal Ward identities and the existence of singular vectors yield
degenerate solutions \cite{Gur93}. With these improved definitions we can
calculate the characters for all irreducible highest weight Jordan cell
representations of the $c_{p,1}$-series. We then write down the modular
invariant partition functions of these models and show that logarithmic
theories may be rational, if one slightly weakens the standard
presumptions for rationality of a CFT. This leads to a new class of
partition functions of CFTs with $c_{{\em eff}} = 1$, which resemble
an {\em A-D-E\/} classification similar to the well known $c=1$ models.
Next, we show that the Verlinde formula for the calculation of the fusion
coefficients from the $S$-matrix cannot be longer valid and discuss some
possible generalizations.
Finally, we discuss the consequences of our results for $c=-2$ to the
theory of polymers, where an additional structure shows up.
  \par
We concentrate ourselves on the case of the $c_{p,1}$ models, since these are
the only ones, where the appearance and behavior of logarithmic operators is
well understood. To fix notation, we shortly review these models here:
  \par
  \subsection{The {\myboldmath $c_{p,1}$} models}
  \pn
In a work of H.G.~Kausch \cite{Kau91} the possibility to extend
the Virasoro algebra by a multiplet of fields of equal conformal
dimension has been considered. Besides some sporadic solutions he found
a series of algebras extended by a singlet or triplet
of fields of odd dimension which resemble a SO(3) structure. The
operator product expansion (OPE) is given by
\begin{equation}\label{eq:walg}
  W^{(j)}(z)W^{(k)}(\zeta) =
    \frac{c}{\Delta}\delta^{jk}\frac{1}{(z-\zeta)^{2\Delta}}
  + C_{\Delta\Delta\Delta} i\varepsilon^{jkl}
    \frac{W^{(l)}(\zeta)}{(z-\zeta)^{\Delta}}
  + {\em descendant\ fields}\,,
\end{equation}
where $c = c_{p,1} = 1-6\frac{(p-1)^2}{p}$ and $\Delta=2p-1$.
%Note, that
%for the singlet algebra there is no term proportional to the field $W$.
These CFT posses infinitely many degenerate representations with integer
conformal weights
\begin{equation}\label{eq:hwertel}
  h_{2k+1,1} = k^2 p + kp - k\,.
\end{equation}
These representations correspond to a set of relatively local chiral
vertex operators. But there is a peculiarity: The energy operator $L_0$
is no longer diagonal on these degenerate representations, but is given
in a Jordan normal form with non-trivial blocks.
  \par
A standard free field construction \cite{BPZ83,DoFa84}
shows that the degenerate fields have conformal weights $h_{m,n} =
\frac{\alpha_{m,n}^2}{4} + \frac{c_{p,1}-1}{24}$, where
$\alpha_{m,n} = m\sqrt{p} - n\sqrt{p}^{-1}$. The fundamental region
of the minimal models unfortunately is empty: $\{m,n|1\leq m<1,\,1\leq n<p\} =
\emptyset$. But without loss of generality we can reduce the
labels $(m,n)$ to the region $0<m, 0<n\leq p$.
%since $\alpha_{m,n} = -\alpha_{-m,-n}$ and $\alpha_{m,n} = \alpha_{m+1,n+p}$.
Moreover, we
have the following abstract fusion rules which result from the
conditions for the existence of well defined chiral vertex operators
\cite{Kau95}.
  \medskip\pano
  {\sc Proposition 1.} {\em Let $c = 13-6(p+p^{-1})$ with $p\in\BN$. Then
  there exist well defined chiral vertex operators for triples of
  Virasoro highest weight representations to $(h_{m_1,n_1}$, $h_{m_2,n_2}$,
  $h_{m_3,n_3})$ with
  $0< m_i$ and $0<n_i\leq p$ iff $|m_1-m_2| <m_3<m_1+m_2$ and
  $|n_1-n_2|<n_3\leq\min(p,n_1+n_2-1)$, and moreover
  $m_1+m_2+m_3-1\equiv n_1+n_2+n_3-1\equiv 0$ {\rm mod} $2$.}
  \medskip\par
The screening charges have a special meaning. With $\alpha_{\pm} =
\alpha_0 \pm \sqrt{1 + \alpha_0^2}$ and $\alpha_0^2 = (1-p)^2/4p$ the
first of them is given by
$$
  Q =\int_{\Omega_1}\frac{dz}{2\pi i}V_{\alpha_+}(z)\,,
$$
where $\Omega_1$ encircles the origin counterclockwise in the standard way.
%and has to be
%chosen in the standard way, if $Q$ is acting on a vertex operator
%$V_{\beta}(w)$.
$Q$ has trivial monodromy on the Fock spaces $\EF_{m,n}$ of the free field
construction on the weights $h_{m,n}$, and therefore is by itself a well
defined local chiral vertex operator
$Q:\EF_{m,n}\rightarrow\EF_{m-2,n}$. This screening charge is
exactly responsible for the multiplet structure of the chiral fields.
We have $Q^m=0$ on $\EF_{m,n}$.
The other screening charge (to the ``power'' $k$) is
$$
  \tilde{Q}^k = \int_{\Omega_k}\frac{dz_1}{2\pi i}\ldots\frac{dz_k}{2\pi i}
  V_{\alpha_-}(z_1)\ldots V_{\alpha_-}(z_k)\,,
$$
where the integration path is radially ordered, $|z_1|> \ldots >|z_k|$,
and encircles the origin. It is well defined on $\EF_{m,n}$ iff
$0<k=n<p$. $\tilde{Q}^p$ vanishes identically on $\EF_{m,p}$.
The BRST-identity is $\tilde{Q}^{p-n}\tilde{Q}^n = 0$, such that we have
the following embedding structure of Fock spaces (see
\cite{Fel89,FFK89}) induced by the exact sequence
$$
  \ldots\stackrel{\tilde{Q}^{p-n}}{\longrightarrow}\EF_{m-2,n}
        \stackrel{\tilde{Q}^{n}}{\longrightarrow}\EF_{m-1,p-n}
        \stackrel{\tilde{Q}^{p-n}}{\longrightarrow}\EF_{m,n}
        \stackrel{\tilde{Q}^{n}}{\longrightarrow}\EF_{m+1,p-n}
        \stackrel{\tilde{Q}^{p-n}}{\longrightarrow}\EF_{m+2,n}
        \stackrel{\tilde{Q}^{n}}{\longrightarrow}\ldots
$$
The Virasoro modules are then given by $\EH_{m,n} =
{\rm ker}_{\EF_{m,n}}\tilde{Q}^n$.
The fields $\phi_{2k+1,1}\equiv V_{\alpha_{2k+1,1}}$,
$k\in\BN$, all have integer dimensions $h_{2k+1,1} = k^2p+kp-k$, such
that one is tempted to extend the local chiral algebra by them.
Indeed, from proposition 1 follows that the local chiral algebra
generated by only the stress energy tensor and the field $\phi_{3,1}$
closes, since no other fields can contribute to the singular part of the
OPE. The multiplet structure is obtained by repeated application of $Q$,
$W^{(j)} = Q^j\phi_{3,1}$. Indeed, this yields three fields with
SO(3)-structure \cite{Kau91}, and therefore a $\w(2,2p-1,2p-1,2p-1)$-algebra.
With $W = \sum_{j}W^{(j)}$ we get the symmetric singlet
algebra $\w(2,2p-1)$.
  \par
With the BRST-structure given above one can construct exactly $2p$
(regular) representations of the fully extended chiral algebra by taking
into account the multiplets generated by the $Q$-operator\footnote{The
operators $Q$ and
$\tilde{Q}^k$ generate four two-dimensional complexes of the $\EF_{m,n}$,
one for $m$ even and odd respectively, and one for $n=p$ and $n\neq p$
respectively \cite{Kau95}.}.
Formally we can write these $\w$-modules as
\begin{eqnarray}              \label{eq:HWp}
  {\EH}^{\w}_{n,+} = \bigoplus_{j=0}^{\infty}\bigoplus_{m=0}^{2j-1}Q^m
  {\EH}_{2j+1,n}\,,\\      \label{eq:HWm}
  {\EH}^{\w}_{n,-} = \bigoplus_{j=1}^{\infty}\bigoplus_{m=0}^{2j-2}Q^m
  {\EH}_{2j,n}\,,
\end{eqnarray}
with $1\leq n\leq p$. The corresponding conformal weights are $h_{1,n}$ and
$h_{2,n}$ respectively. The $\w$-representations for
$h_{1,n}$ are singlets, the ones for $h_{2,n}$ doublets.
There also exist special representations for the weights $h_{0,n},
1\leq n<p$. Their highest weight vectors are singular vectors in
$\EF_{1,p-n}$, which have the {\em same} highest weights. The
corresponding chiral vertex operators are degenerated. For instance,
there are besides the identity $p-1$ additional vertex operators of
conformal weight zero, which map $\EF_{0,n}$ to
$\EF_{1,p-1}$. Consequently, also the descendant fields of the
identity family are degenerated, in particular the Virasoro field itself.
This forces the existence of non-trivial Jordan cells for $L_0$, i.e.\ $L_0$
no longer is diagonalizable. Moreover, the multiplicities of states in
the Virasoro modules must change. We have, in sloppy terms, a $p$-fold
degenerate identity, which will lead to a multiplicity of $p$ in the
characters of the highest weight representations $h_{0,n}$.
  \par
  \subsection{Structure constants}
  \pn
Although the chiral vertex operator algebra is degenerated, it is
possible to explicitly calculate their structure constants with the
methods given in \cite{FFK89,Flo93}.
First, we make the usual ansatz for the decomposition of local chiral
fields into chiral vertex operators,
\begin{equation}\label{eq:cvo-decomp}
  W^{(n,n')}(z) = \sum_{l,l',m,m'}{\cal D}^{(l,l')}_{(n,n')(m,m')}
  V_{h_{l,l'}h_{n,n'}}^{h_{m,m'}}(\cdot,z)\,,
\end{equation}
where the chiral vertex operators are maps $\EH_{h_{n,n'}} \mapsto
{\rm Hom}(\EH_{h_{m,m'}},\EH_{h_{l,l'}})$. In our case we have
$n' = m' = l' = 1$ and $n, m, l$ odd, where $h_{2k+1,1}$ is given by
(\ref{eq:hwertel}). The coefficients
${\cal D}^{(l,1)}_{(n,1)(m,1)}$ important for us are given by
\begin{eqnarray}\label{eq:d-coeff}
  {\ds\left({\cal D}_{(n,1)(m,1)}^{(l,1)}\right)^2} &=&
  % {\ds c\cdot\frac{h_{l,1}}{h_{n,1}h_{m,1}}D_{(n,1)(m,1)}^{(l,1)}\ =\
  {\ds c\cdot\frac{h_{l,1}}{h_{n,1}h_{m,1}}
  \frac{N_{(l,1)(l,1)}^{(1,1)}}{N_{(n,1)(n,1)}^{(1,1)}N_{(m,1)(m,1)}^{(1,1)}}
  \left(\Delta_{n,m}^{l}(x)\Delta_{1,1}^{1}(x')\right)^2}\,,\\
  {\ds\Delta_{n,m}^{l}(x)} &=& {\ds (-1)^{\frac{1}{2}(n+m-l-1)}
  \left(\frac{[n]_x[m]_x[l]_x}{[1]_x}\right)^{\frac{1}{2}}}\nonumber\\*
  &\times&{\ds\prod_{j=(l+n-m+1)/2}^{n-1}[j]_x
  \prod_{j=(m+n-l+1)/2}^{n-1}[j]_x
  \prod_{j=(l+m-n+1)/2}^{(l+m+n-1)/2}\frac{1}{[j]_x}}\,.
\end{eqnarray}
Here, we have defined $[j]_x = x^{j/2} - x^{-j/2}$ with $x = \exp(2\pi ip)$
and $x' = \exp(2\pi ip^{-1})$.
The prefactor $c\cdot h_{l,1}h_{n,1}^{-1}h_{m,1}^{-1}$ stems from our
normalization of the two-point-functions, which we have chosen for
simple primary fields to be
\begin{equation}\label{eq:tpoint}
  \avac{0}W^{(n,1)}_{-h_{n,1}}W^{(m,1)}_{h_{m,1}}\vac{0} =
  \frac{c}{h_{n,1}}\delta_{n,m}\,.
\end{equation}
The  normalization integrals $N_{(n,1)(m,1)}^{(l,1)}$, i.e.\ the
three-point-functions of the chiral vertex operators, can be derived from
the general solution of the Fuchsian differential equations for
degenerate representations of the Virasoro algebra \cite{DoFa84}.
They are particularly simple in our case. With $r = \frac{1}{2}(n+m-l-1)$
denoting the number of inserted screening charges they read
\begin{eqnarray}\label{eq:norm}
  {\ds N_{(n,1)(m,1)}^{(l,1)}} &=&
  {\ds (-1)^{\frac{1}{2}r}
  \prod_{j=1}^{r}\frac{[m-j]_x[j]_x}{[1]_x}}\nonumber\\*
  &\times&
  {\ds\prod_{j=1}^{r}\frac{\Gamma(jp)
  \Gamma(1+(j-m)p)\Gamma(1+(j-n)p)}{\Gamma(p)\Gamma(2+(r-m-n+j)p)}}\,.
\end{eqnarray}
The structure constants of the OPE or equivalently of the Lie algebra of
the Fourier modes of the local chiral fields are then
\begin{equation}\label{eq:strconst}
  C_{(n,1)(m,1)}^{(l,1)} =
  {\cal D}_{(n,1)(m,1)}^{(l,1)}N_{(n,1)(m,1)}^{(l,1)}\,.
\end{equation}
Let us introduce the following symmetry factor
${\cal S}(n,m,l)$, since the extended symmetry algebra, due to its
{\euf su}(2)-structure, contains either a multiplet of fields of the
same dimension
or the symmetric singlet of the latter. The reason for this is that the
screening charge itself is a local operator acting on local fields, as
long as $p\in\BN$. In our case we have
multiplets $W^{(j)}_l = Q^jW^{(l,1)}$, $0\leq j < l$, and their symmetric
singlets $W_l = \sum_{j=0}^{l-1} W^{(j)}_l$. From the formal fusion
rules for degenerated representations
\begin{equation}\label{eq:fus1}
  W^{(j)}_l \star W^{(j')}_{l'} =
  \sum_{{m=|l-l'|+1 \atop l+l'-m-1\,\equiv\,0\,{\rm mod}\,2}}^{l+l'-1}
  W^{(j+j'-r)}_m\,,
\end{equation}
where again $r = \frac{1}{2}(l+l'-m-1)$, we can read off the
multiplicities for the formal fusion rules of the singlets,
\begin{equation}\label{eq:fus2}
  W_l \star W_{l'} =
  \sum_{{m=|l-l'|+1 \atop l+l'-m-1\,\equiv\,0\,{\rm mod}\,2}}^{l+l'-1}
  {\cal S}(l,l',m)W_m =
  \sum_{{m=|l-l'|+1 \atop l+l'-m-1\,\equiv\,0\,{\rm mod}\,2}}^{l+l'-1}
  {\frac{1}{2}(l+l'-|l-l'|-2)\choose\frac{1}{2}(m-|l-l'|-1)}W_m\,,
\end{equation}
which essentially are the multiplicities arising in tensoring
symmetric {\euf su}(2) Young tableaus.
Next, we determine the phases $\Delta_{n,m}^l(x)$. Trivially,
the phase $\Delta_{1,1}^1(x') = 1$ for all $x'$.
Since $p\in\BN$, we find that
$[k]_x=0$. Actually, considering $[k]_x$ for $p + \eps$ in the limit
$\eps\rightarrow 0$, we get to leading order of $\eps$ the expression
\begin{equation}\label{eq:q-symbols}
  [k]_x = (-1)^{kp}2\pi ik\eps\ \ \ (\eps\rightarrow 0)
\end{equation}
for $p\in\BN$. All these $2\pi i\eps$ factors exactly cancel in
numerator and denominator such that $\Delta_{n,m}^l(x)$ is analytic also for
integer $p$. Naively, one would expect that the result is just a sign,
since all considered operators are local to each other. Instead of this
we find that
$\left(\Delta_{n,m}^l(x)\right)^2$ are rational numbers, which stem from
the multiplet structure,
\begin{eqnarray}\label{eq:phase}
  \Delta_{n,m}^l(x) &=& (-1)^{lp}(-1)^{\frac{1}{2}(n+m-l-1)(p+1)}
  (-1)^{p((lm+ln+nm)/2-(l^2+m^2+n^2-1)/4)}\nonumber\\*
  &\times&\frac{\sqrt{nml}(n-1)!^2(\frac{1}{2}(l+m-n-1))!}{
  (\frac{1}{2}(l+n-m-1))!(\frac{1}{2}(m+n-l-1))!(\frac{1}{2}(l+m+n-1))!}\,.
\end{eqnarray}
In the same way we get singularities in the normalization integrals for
$p\in\BN$. Nonetheless, the square of $C_{(n,1)(m,1)}^{(l,1)}$ is still
well defined and finite, since all singularities cancel.
Moreover, the values for integer $p$ are precisely given as the
analytical continuation of the region $p\in\BR_{+}-\BN$:
In the limit $\eps\rightarrow 0$ we get for $m\in\BZ$, $p\in\BN$ to leading
order in $\eps$
\begin{equation}\label{eq:gamma}
  \Gamma(m(p+\eps)) = \left\{\begin{array}{cl}
    (mp - 1)! & \ \ mp > 0\,,\\
    \frac{(-1)^{mp}}{(-mp)!}\frac{1}{m\eps} & \ \ mp\leq 0\,.
  \end{array}\right.
\end{equation}
Finally, plugging this in the normalization integrals, we arrive at the result
\begin{eqnarray}\label{eq:cijk}
  \left(C_{(n,1)(m,1)}^{(l,1)}\right)^2 &=& \frac{c}{{\cal S}(n,m,l)}
  \frac{h_{l,1}}{h_{m,1}h_{n,1}}(\varphi(\Delta_{n,m}^{l}))^2
  \prod_{j=1}^{r}
  \frac{((pj-1)!^2(p(r+l+1-j)-2)!^2}{p(m-j)-1)!^2(p(n-j)-1)!^2}\nonumber\\*
  &\times&
  \prod_{j=1}^{n-1}\frac{(p(n-j)-1)!^2}{(pj-1)!(p(n-j+1)-2)!}
  \prod_{j=1}^{m-1}\frac{(p(m-j)-1)!^2}{(pj-1)!(p(m-j+1)-2)!}\nonumber\\*
  &\times&
  \prod_{j=1}^{l-1}\frac{(pj-1)!(p(l-j+1)-2)!}{(p(l-j)-1)!^2}\,,
\end{eqnarray}
where again $r = \frac{1}{2}(n+m-l-1)$. We denote by
$\varphi(\Delta_{n,m}^{l})$ the phase part (a sign) of
$\Delta_{n,m}^{l}$ from (\ref{eq:phase}),
since the modulus combines nicely with corresponding terms from the
normalization integrals to ${\cal S}(n,m,l)^{-1}$.
  \par
We may check these formulas with already known results. Firstly, we
obtain the following explicit expression for the self-coupling constant
$C_{\Delta\Delta}^{\Delta}$ in (\ref{eq:walg}),
\begin{equation}\label{eq:c333}
  \left(C_{(3,1)(3,1)}^{(3,1)}\right)^2 = c_{p,1}(-1)^{p}
  \frac{(4p-2)!^2(p-1)!^3}{2(3p-2)!(2p-1)!^4}\,.
\end{equation}
Here, the {\euf su}(2) multiplicity of the triplet on the rhs of a
symmetric tensor product of two triplets is ${\cal S}(3,3,3)=2$.
The expression (\ref{eq:c333}) is precisely the one earlier obtained by
H.G.~Kausch \cite{Kau91} by explicitly integrating the screening charges.
Since the field $W_5$ has even dimension $h_{5,1} = 6p - 2$, its OPE
with itself has no term proportional to $W_3$ with the odd dimension
$h_{3,1} = 2p-1$. Therefore, we may construct a non-trivial, even sector
subalgebra ${\cal W}(2,6p-2)$. In fact, the next field of even dimension,
$W_{9}$, has $h_{9,1} = 20p - 4 > 2h_{5,1} - 1$, and thus does not
appear on the rhs of the OPE. Since $h_{5,1}$ is even, the self coupling
does not vanish even in the singlet case. Two examples of this
subalgebra series could be constructed explicitly, namely
${\cal W}(2,4)$ at $c=1$ \cite{BFKNRV91,KaWa91} and ${\cal W}(2,10)$
at $c = -2$ \cite{EHH93b}. The self-coupling constants are
$\frac{50}{3}$ and $-\frac{352836}{5}$ respectively. Our explicit
formula (\ref{eq:cijk}) yields with ${\cal S}(5,5,5) = 6$ the expression
\begin{equation}\label{eq:c555}
  \left(C_{(5,1)(5,1)}^{(5,1)}\right)^2 = c_{p,1}(-1)^{p-1}
  \frac{(2p-1)!^3(p-1)!^3(7p-2)!^2(6p-2)!(6p-3)!}{6(4p-1)!^3(3p-1)!^3
  (5p-2)!(4p-2)!(3p-2)!(2p-2)!}\,,
\end{equation}
which for $p=1$ and $p=2$ gives the desired numbers.
%
%%< REPRESENTATIONS AND CHARACTERS >%%%%%%%%%%%%%%%%%%%%%%%%%%%%%%%%%%%%%%
%
  \par
  \mysection{Representations and Characters}
  \pn
Let us assume that the Hilbert space $\EH\otimes\bar{\EH}$ is a
direct sum of irreducible highest weight representations (HWR) with
respect to the chiral symmetry algebra $\w$,
\begin{equation}
  \EH\otimes\bar{\EH} = \bigoplus_{\lambda\in\Lambda}
  \EH^{(\lambda)}
  \otimes\bigoplus_{\bar{\lambda}\in\bar{\Lambda}}
  \bar{\EH}^{(\bar{\lambda})}\,.
\end{equation}
Further we assume that $\w$ is maximal such that $\Lambda =
\bar{\Lambda}$ is the set of all $\w$ HWRs, i.e.\
the theory is {\em symmetric}. We decompose $\EH^{(\lambda)}$ into
Virasoro HWRs, the set of them we denote with $N_{\lambda}$,
\begin{equation}\label{eq:WDecomp}
  \EH\otimes\bar{\EH} = \bigoplus_{\lambda\in \Lambda}\left(
  \bigoplus_{\nu\in N_{\lambda}}\EH^{(\lambda)}_{\nu}
  \otimes\bigoplus_{\nu\in N_{\lambda}}
  \bar{\EH}^{(\lambda)}_{\nu}\right)\,.
\end{equation}
A CFT is said to be {\em rational}, iff $|\Lambda|<\infty$. It is called
{\em quasi-rational}, if $\Lambda$ is countable.
The Cartan subalgebra $\cal C$ is spanned by $L_0$, the central
extension $C$ and the zero modes of the simple primary fields
$\phi_i\in{\cal B}_{{\cal W}}$ which generate the $\w$-algebra.
We denote the highest weight state (HWS) of $\EH^{(\lambda)}_{\nu}$ by
$\vak{h^{(\lambda)}} = \vac{c,h_{\nu},w^{(\lambda)}_1,w^{(\lambda)}_2,\ldots}$
where ${\bf h}\in{\cal C}^*$ is the highest weight vector (HWV).
A regular HWR $M_{\vak{h}}$ of a $\w$-algebra to a HWS
$\vak{h} = \vac{c,h,w_1,w_2,\ldots}$ is then defined to satisfy the
following conditions:
  \begin{eqnarray*}
    C\vak{h} & = & c\vak{h}\,,\nonumber\\
    L_0\vak{h} & = & h\vak{h}\,,\nonumber\\
    \phi_{i,0}\vak{h} & = & w_i\vak{h}\ \ \forall\phi_i\in{\cal B}_{\w}\,,
    \nonumber\\
    L_n\vak{h} & = & 0\ \ \forall n>0\,,\nonumber\\
    \phi_{i,n}\vak{h} & = & 0\ \ \forall\phi_i\in{\cal B}_{\w}\
    {\rm and}\ \forall n>0\,,\nonumber\\
    M_{\vak{h}} & = & U(\w)\vak{h}\,,
  \end{eqnarray*}
where $U(\w)$ denotes the universal enveloping algebra of $\w$. Moreover,
we call a HWR $V_{\vak{h}}$ {\em Verma module}, iff the sequence
  \begin{equation}
    V_{\vak{h}} \longrightarrow M_{\vak{h}} \longrightarrow 0
  \end{equation}
is exact for all HWRs $M_{\vak{h}}$.
The Verma module $V_{\vak{h}}$ has a natural gradation
\begin{equation}
  V_{\vak{h}} = \bigoplus_{n\in\BZ_+}V_{\vak{h}}^n\,,
\end{equation}
where $V_{\vak{h}}^n$ is the $L_0$ eigenspace with eigenvalue $h + n$.
  \par
Let us now assume that there exist HWRs, whose $L_0$ eigenvalues differ
by integers. We must distinguish two cases. If the difference $\Delta h$
of the $L_0$ eigenvalues of two HWRs is always non zero, or the HWVs
differ in at least one component, it still is
possible to diagonalize $L_0$, even if $\Delta h\in\BZ$. Moreover, there
are no logarithmic operators necessary. The reason is that the differential
equations for the conformal Ward identities do not degenerate in this
case. This is different to the case of the modular differential equation
to be satisfied by the characters, which is only sensible modulo integers.
Examples of such rational CFTs with HWRs with $\Delta h\in\BZ$ can
be found in \cite{Flo93,Flo94}.
  \par
Therefore, we now assume the existence of $n+1>1$ HWRs such that
${\bf h}_i-{\bf h}_j=0$ for $1\leq i,j\leq n+1$.
We call such CFTs {\em logarithmic}.
As V.~Gurarie \cite{Gur93} pointed out, we have to modify the definition
of HWRs in the following way: The HWS is replaced by a non-trivial
Jordan cell of $L_0$ of dimension $n+1$, which is spanned by
$\{\vac{{\bf h};0}=\vak{h},\vac{{\bf h};1},\ldots,\vac{{\bf h};n}\}$.
We then will call $M(\vak{h;m})_{0\leq m\leq n}$ a {\em logarithmic\/} HWR
of a $\w$-algebra to the highest weight $L_0$-Jordan cell of rank $n+1$,
$(\vak{h;m} = \vac{c,h,w_1,w_2,\ldots;m})_{0\leq m\leq n}$, if
it satisfies the following conditions:
  \begin{equation}
    \begin{array}{rcl}
      L_0\vac{{\bf h};m} & = & h\vac{{\bf h};m} + \vac{{\bf h};m-1}\,,
      \ \ m>0\,,\\
      L_0\vac{{\bf h};0} & = & h\vac{{\bf h};0}\,,\\
      \phi_{i,0}\vac{{\bf h};m} & = & w_i\vac{{\bf h};m} + \ldots\,,
      \ \ m>0\,,\ \ \forall\phi_i\in{\cal B}_{\w}\,,
    \end{array}
  \end{equation}
and otherwise the conditions of the original definition. The dots in the
last condition represent possible non-diagonal contributions.
In addition, there is in general no orthogonal system of states within the
Jordan cell, i.e.\ $\vev{{\bf h};k}{{\bf h};l}\neq 0$ even for $k\neq l$.
  \par
Since the other properties of HWRs remain unchanged, it makes sense to
consider such logarithmic HWRs if the whole Jordan cell structure is taken
into account for the definition of $\w$-families.
  \par
Next, we want to discuss the consequences for the characters.
For simplicity, we consider a Jordan cell of form
${h\ 1\choose 0\ h}$, i.e.\ we have two HWSs,
$\vac{h;0}$ and $\vac{h;1}$, on which the action of $L_0$ is given by
$L_0\vac{h;0} = h\vac{h;0}$ and $L_0\vac{h;1} = h\vac{h;1} + \vac{h;0}$.
The off-diagonal element could be any non-zero number, since a Jordan
cell decomposition is just one particular choice. The physical correct
decomposition will be fixed later by modular invariance.
  \par
The HWS $\vac{h;0}$ is an ordinary $L_0$-eigenstate, such that the
character of the corresponding HWR should be defined in the usual manner.
The other state, $\vac{h;1}$ is not a $L_0$-eigenstate, application of
$L_0$ generates a new state, which is not contained in the standard
Verma module. If we apply $L_0$ once again, this state is recovered plus
an additional one, etc. Thus, the operator $L_0$, acting on the Jordan
cell, may be written as
$L_0 = {L_{0;0}\ \ 1\choose 0\ \ L_{0;1}}$, where the second label $j$
refers to the Verma like modules on which the $L_{0;j}$ operators act.
  \par
The character of a HWR on a HWS $\vak{h}$ is usually defined as
\be\label{eq:trace}
  \chi_{\vak{h}}(q) = {\rm tr}_{M_{\vak{h}}}q^{L_0-c/24}
  % = \sum_{\psi\in {\cal B}(M_{\vak{h}})}\vev{\psi}{q^{L_0-c/24}\psi}
  \,,
\ee
where $q=\exp(2\pi i\tau)$ is up to now a formal variable, and the trace
is taken over the module which is created by action of $U({\cal W})$ on
$\vak{h}$. Using our $L_0$ matrix, we obtain
\begin{eqnarray}
  q^{L_0} & = & \sum_{n=0}^{\infty} \frac{(2\pi i\tau)^n}{n!}\left(
  \begin{array}{cc} L_{0;0} & 1\\
                          0 & L_{0;1}
  \end{array}\right)^n\nonumber\\
          & = & \sum_{n=0}^{\infty} \frac{(2\pi i\tau)^n}{n!}\left(
  \begin{array}{cc} L_{0;0}^n & nL_{0;0}^{n-1}\\
                            0 & L_{0;1}^n
  \end{array}\right)^n\nonumber\\
          & = & \left(
  \begin{array}{cc} q^{L_{0;0}} & 2\pi i\tau q^{L_{0;0}}\\
                              0 & q^{L_{0;1}}
  \end{array}\right)\,.
\end{eqnarray}
Since formally $2\pi i\tau = \log(q)$, we see that a non-trivial Jordan
cell may generate logarithmic terms in the character expansions. This is
completely analogous to the logarithms in the correlation functions of
certain operators, which stem from the degeneracy of the conformal Ward
identity differential equations: We obtain essentially the same
degeneracies in the modular differential equations for the characters,
which force additional solutions with logarithms.
  \par
The careful reader may wonder, how the logarithmic terms can show up in the
characters. Usually, traces (\ref{eq:trace}) over modules are well defined,
since the complete Hilbert space is a direct sum of modules and
$L_0$ can be uniquely restricted to one of the modules.
Now, if $L_0$ has non trivial Jordan form, modules
$M_{\vac{{\bf h};k}}$ and $M_{\vac{{\bf h};l}}$ are not orthogonal.
Therefore, the characters depend on the choice of a basis of generating
states, while the sum $\sum_{k=0}^n \chi_{\vac{{\bf h};k}}(q)$ is invariant
under any basis change $\vac{{\bf \tilde h};k} = B^k_{\phantom{k}l}
\vac{{\bf h};l}$.
Only this sum is a trace of a well defined restriction of $q^{L_0-c/24}$
and does never contain any logarithmic parts. But the characters can:
For example change of the basis $\{\vac{h;0},\vac{h;1}\}$ to
$\{\vac{\tilde h;0}=\vac{h;0}+\vac{h;1},
\vac{\tilde h;0}=-\vac{h;0}+\vac{h;1}\}$ yields
$$
  q^{L_0} = \frac{1}{2}\left(
  \begin{array}{cc}
    (1-2\pi i\tau)q^{L_{0;0}}+q^{L_{0;1}} &
    (1+2\pi i\tau)q^{L_{0;0}}-q^{L_{0;1}}\\
    (1-2\pi i\tau)q^{L_{0;0}}-q^{L_{0;1}} &
    (1+2\pi i\tau)q^{L_{0;0}}+q^{L_{0;1}}
  \end{array}\right)\,.
$$
We will return to this point in the last section,
where an explicit realization of this is given for the CFT at $c=-2$.
%There we have two HWSs with conformal weight $h=0$, $\vac{0}$ and
%$\vac{\xi}=\xi_0\vac{0}$, which span a Jordan cell such that $L_0$ has
%Jordan form.
The generalization to larger Jordan cells is straightforward.
  \par
Since the characters of a CFT can be viewed as the zero-point-functions
on a torus with modular parameter $\tau$, they in general turn out to be
certain modular functions whose Fourier expansions around $\tau=i\infty$
are just the $q$-series. One of the most powerful tools in CFT is the
modular invariance of the partition function
\begin{equation}\label{eq:PartFct}
  Z(\tau,\bar{\tau}) = (q\bar q)^{\frac{c}{24}}
  {\rm tr}(q^{L_0}\bar q^{\bar L_0})\,.
\end{equation}
The exciting result of J.L.~Cardy \cite{Car86}, which was then
mathematical rigorously proven by W.~Nahm \cite{Nah91}, is that
conformal invariance of a field theory on $S^2$ implies modular
invariance of the partition function of the field theory on a torus.
Unfortunately, the proof is strictly valid only for theories with
diagonalizable $L_0$, but the result should also be valid for
logarithmic CFTs (if the full Jordan cell is taken as basis for the
HWRs), since the logarithmic behavior appears as a consequence of
degeneracies of certain differential equations due to discrete and
isolated values in the parameter space (essentially ${\bf h}\in{\cal
C}^*$). Therefore, we expect that modular invariance can be extended to
the logarithmic case in the same way as the logarithmic solutions of the
differential equations are analytic continuations from the regular case.
Since the partition function is a
quadratic form in the characters, modular invariance puts severe
restrictions on the modular behavior of the (generalized) characters.
It will turn out that modular invariance uniquely determines a basis
of HWSs within each Jordan block and therefore all characters.
  \par
We now fix some notations for the following.
We will very often use the so called {\em elliptic functions\/} or
{\em Jacobi-Riemann $\Theta$-functions\/} which are modular forms of
weight $1/2$, defined as
\begin{equation}\label{eq:thetadef}
  \Theta_{\lambda,k}(\tau) = \sum_{n\in\BZ}q^{(2kn + \lambda)^2/4k}\,,\ \
  \lambda \in \BZ/2\,,\ k \in \BN/2\,.
\end{equation}
We call $\lambda$ the {\em index\/} and $k$ the {\em modulus\/} of the
$\Theta$-function. The $\Theta$-functions obey
$\Theta_{\lambda,k} = \Theta_{-\lambda,k}
= \Theta_{\lambda+2k,k}$, and $\Theta_{k,k}$ has, as power series in
$q$, only even coefficients. We also need the {\em Dedekind
$\eta$-function\/} which is defined as $\eta(\tau) =
q^{1/24}\prod_{n\in\BN}(1-q^n)$. The modular properties of these
functions are for $\lambda,k\in\BZ$
\begin{eqnarray}\label{eq:theta}
  \Theta_{\lambda,k}({\ts-\frac{1}{\tau}}) &=& {\ds
  {\ts\sqrt{\frac{-i\tau}{2k}}}\,\sum_{\lambda'=0}^{2k-1}
  e^{i\pi\frac{\lambda\lambda'}{k}}\Theta_{\lambda',k}(\tau)}\,,\\
  \Theta_{\lambda,k}({\ts\tau + 1}) &=& {\ds
  e^{i\pi\frac{\lambda^2}{2k}}\Theta_{\lambda,k}(\tau)}\,,\\
  \eta({\ts-\frac{1}{\tau}}) &=& {\ds\sqrt{-i\tau}\,\eta(\tau)}\,,\\
  \eta({\ts\tau + 1})        &=& {\ds e^{\pi i/12}\,\eta(\tau)}\,.
\end{eqnarray}
The functions $\Lambda_{\lambda,k}(\tau) = \Theta_{\lambda,k}(\tau)/\eta(\tau)$
are then modular forms of weight zero to a particular main-congruence
subgroup $\Gamma(N)\subset{\rm PSL}(2,\BZ)$, e.g.\ $N$ is the least
common multiple of $4k$ and $24$ for $k\in\BZ$.
  \par
As we have seen above, the characters for logarithmic CFTs are functions
in the ring $\BZ[[q]][\log q]$. Therefore we introduce the following
additional functions:
\be
  (\partial\Theta)_{\lambda,k}(\tau) \propto
  \frac{\partial}{\partial\lambda}\Theta_{\lambda,k}(\tau) =
  \frac{2\pi i\tau}{k}\sum_{n\in\BZ}(2kn+\lambda)q^{(2kn + \lambda)^2/4k}\,,
\ee
where we made explicit that new linear independent solutions of
degenerate differential equations can be obtained by a formal derivation
of the degenerate solution with respect to its parameter. As long as
modular covariance is not concerned, there is no reason why $\tau$ could
not appear as a factor. We introduce the so-called
{\em affine\/} $\Theta$-functions
\begin{equation}
  (\partial\Theta)_{\lambda,k}(\tau) =
  \sum_{n\in\BZ}(2kn+\lambda)q^{(2kn + \lambda)^2/4k}\,,
\end{equation}
which play an important r{\^o}le in the character formulas for the affine
$\widehat{{\euf su}(2)}$-algebra. They are odd,
i.e.\ $(\partial\Theta)_{-\lambda,k} =
-(\partial\Theta)_{\lambda,k}$. Moreover, per definitionem
$(\partial\Theta)_{0,k} = (\partial\Theta)_{k,k} \equiv 0$.
Their modular behavior is
\begin{equation}
  \begin{array}{rcl}
    (\partial\Theta)_{\lambda,k}({\ts-\frac{1}{\tau}}) &=& {\ds
    {\ts(-i\tau)\sqrt{\frac{-i\tau}{2k}}}\,\sum_{\lambda'=1}^{2k-1}
    e^{i\pi\frac{\lambda\lambda'}{k}}
    (\partial\Theta)_{\lambda',k}(\tau)}\,,\\
    (\partial\Theta)_{\lambda,k}({\ts\tau + 1}) &=& {\ds
    e^{i\pi\frac{\lambda^2}{2k}}
    (\partial\Theta)_{\lambda,k}(\tau)}\,.\\
  \end{array}
\end{equation}
Since they are no longer modular forms of weight $1/2$ under
$S:\tau\mapsto -1/\tau$, we have to add further functions
\begin{equation}
  (\nabla\Theta)_{\lambda,k}(\tau) =
  \frac{\log q}{2\pi i}\sum_{n\in\BZ}(2kn+\lambda)q^{(2kn + \lambda)^2/4k}
\end{equation}
in order to obtain a closed finite dimensional representation of the
modular group. It is clear that $S$ interchanges these two sets of
functions, while $T:\tau\mapsto\tau+1$ transforms $(\nabla\Theta)_{\lambda,k}$
into $(\nabla\Theta)_{\lambda,k}+
(\partial\Theta)_{\lambda,k}$. Therefore, the linear combination
$$
  (\partial\Theta)_{\lambda,k}(\tau)(\nabla\Theta)_{\lambda,k}^*(\bar{\tau}) -
  (\nabla\Theta)_{\lambda,k}(\tau)(\partial\Theta)_{\lambda,k}^*(\bar{\tau}) =
  (\tau - \bar{\tau})|(\partial\Theta)_{\lambda,k}|^2
$$
is modular covariant of weight 1/2!
  \par
Of course, the modular differential equation could be degenerate of
higher degree, and one had to introduce generalizations
$(\partial^n\Theta)_{\lambda,k}$ and
$(\nabla^n\Theta)_{\lambda,k}$ (the expression $(\tau - \bar{\tau})^n$
is modular covariant of weight $-2n$ for all $n\in\BZ_+$).
One can show \cite{Eho00} that regular rational theories with
$c_{{\em eff}} \leq 1$ can only have one power $\eta(\tau)\eta(\bar{\tau})$
in the denominator of the partition function. Regular means that the characters
are modular forms.
Now, the modular behavior of characters of logarithmic CFTs
is almost the one of modular forms, except the possibility to
expand into a power series in $q$. In particular, the asymptotic properties
needed in the proof \cite{Eho00} are only affected in an analytic way by
logarithmic corrections: In fact, although the modular differential equation
makes only sense for particular isolated points in parameter space,
$(c,{\bf h}_1,{\bf h}_2,\ldots)\in\bigoplus{\cal C}^*$,
where the corresponding CFT is rational, it can be regarded as a
differential equation depending on continuously variable parameters --
once it has been written down. The characters of our theories in question
are solutions of certain degenerate modular differential equations,
obtained in a unique way by analytic continuation.
Therefore, we conjecture that the result of \cite{Eho00} should also hold
for logarithmic RCFTs. Thus, we should only be concerned with $n=1$ in our
case.
%
%%< CHARACTERS FOR THE $c_{p,1}$ MODELS >%%%%%%%%%%%%%%%%%%%%%%%%%%%%%%%%%
%
  \par
  \mysection{Characters for the {\myboldmath $c_{p,1}$} models}
  \pn
We already have seen that the $c_{p,1}$ models are logarithmic CFTs.
V.~Gurarie \cite{Gur93} first derived the consequences of the existence
of logarithmic fields. He explicitly considered the example of the
degenerate Virasoro model at $c=-2$, i.e.\ the $c_{2,1}$ model. As
explained above, logarithmic CFTs always have {\em inequivalent} HWRs to the
same HWV {\bf h}. A more detailed analysis of degenerate models shows
that then always a further representation exists which has the generic
null vector of the (regular) HWR on $\vak{h}$ as HWS. This reflects the
fact that many properties of the characters are only defined modulo
$\BZ$, since the HWR on the null vector of the representation
$\EH_{\vac{h}}$ has the highest weight $h+k$ with $k\in\BZ_+$.
We are now going to derive the characters of these models. First, we show that
the singlet models $\w(2,2p-1)$ are not rational since the chiral
symmetry algebra is too small for that.
  \par
  \subsection{Characters of the singlet algebras {\myboldmath $\w(2,2p-1)$}}
  \pn
The additional primary field of the $\w(2,2p-1)$-algebra is just the
symmetric singlet of the {\euf su}(2) triplet of primary fields which
generate the $\w(2,2p-1,2p-1,2p-1)$. One way to obtain the characters is
to explicitly calculate the vacuum character and then get the others by
modular transformations. From the embedding structure of Virasoro Verma
modules for the values $c=c_{p,1}$ of the central charge
\cite{FeFu82,FeFu83,Fel89,FFK89} we learn that the Virasoro character
for the HWR on $\vac{h_{2n+1,1}}$, $n\in\BZ_+$, is given by
\begin{equation}
  \chi^{{\em Vir}}_{2n+1,1}(\tau) = \frac{q^{(1-c)/24}}{\eta(\tau)}\left(
  q^{h_{2n+1,1}} - q^{h_{-2n-1,1}}\right)\,.
\end{equation}
Therefore \cite{Flo93}, the character of the ${\cal W}$-algebra vacuum
representation is
\begin{eqnarray}
  \chi^{{\cal W}}_{0}(\tau)&=&\sum_{n\in\BZ_+}\chi^{{\em Vir}}_{2n+1,1}(\tau)\\
                           &=&\frac{q^{(1-c)/24}}{\eta(\tau)}
             \sum_{n\in\BZ}{\rm sgn}(n)q^{\frac{(2pn+p-1)^2}{4p}}\,,
\end{eqnarray}
where we defined ${\rm sgn}(0) = 0$.
It is convenient to rewrite the signum function as
${\rm sgn}(n+\frac{p-1}{2p})$. This character seems (up to the signum
function) to be quite similar to the classical ${\euf su}(2)$-$\Theta$-function
$\Theta_{p-1,p}(\tau,0,0)$ divided by $\eta$.
Note, that the classical ${\euf su}(2)$-$\Theta$-functions
$\Theta_{\lambda,k}(\tau,z,u)$, coincide for
$z=u=0$ with the elliptic functions defined in (\ref{eq:thetadef}).
They are the building stones for the characters of the
$\widehat{{\euf su}(2)}$ Kac-Moody-algebra. We therefore define
\begin{equation}
  \Xi_{n,m}(\tau) = \sum_{k\in\BZ+\frac{n}{2m}}{\rm sgn}(k)q^{mk^2}\,.
\end{equation}
But the modular transformation behavior is quite different from
(\ref{eq:theta}), while the presence of the signum function does not
change the behavior under $T$,
$\Xi_{n,m}(\tau+1) = \exp(i\pi\frac{n^2}{2m})\Xi_{n,m}(\tau)$.
In order to get the behavior under $S$, we rewrite the functions
$\Xi_{n,m}$ as linear combinations of
$\Theta_{\lambda,k}$ functions. For this we introduce
\begin{equation}
  \sigma(x,y) = \lim_{\varepsilon\rightarrow 0}\frac{1}{\sqrt{2\pi}}
  \int_{-\infty}^{\infty}\frac{e^{-2\pi iyp^2}}{p+i\varepsilon^2}\left(
  e^{ipx} - e^{-ipx}\right)dp\,,
\end{equation}
such that $\sigma(x,0) = {\rm sgn}(x)$. In the following we omit the
obvious limiting procedure. We find
\begin{eqnarray}
  \Xi_{n,m}(\tau) &=& \sum_{k\in\BZ+\frac{n}{2m}}\sigma(k,0)q^{mk^2}\nonumber\\
                  &=& \sum_{k\in\BZ+\frac{n}{2m}}\frac{1}{\sqrt{2\pi}}
                      \int_{-\infty}^{\infty}\frac{dp}{p}\left(
                      e^{2\pi ikp} - e^{-2\pi ikp}\right)
                      q^{mk^2}\\
                  &=& \frac{1}{\sqrt{2\pi}}\int_{-\infty}^{\infty}
                      \frac{dp}{p}\left(
                      \Theta_{n,m}(\tau,p,0)-\Theta_{n,m}(\tau,-p,0)\right)
                      \nonumber\,.
\end{eqnarray}
Therefore, by linearity of the $S$-transformation, we can write
\begin{equation}
  \Xi_{n,m}(-\frac{1}{\tau}) = \widetilde{\Xi}_{n,m}(\tau) =
  \sqrt{\frac{-i\tau}{2m}}
  \sum_{n'\,{\rm mod}\,2m}\sin(-2\pi\frac{nn'}{2m})\Xi_{n',m}(\tau)\,,
\end{equation}
where $\widetilde{\Xi}_{n,m}$ is given by
\begin{eqnarray}
  \widetilde{\Xi}_{n,m} &=& \sum_{k\in\BZ+\frac{n}{2m}}
                            \sigma(k,-\frac{1}{2m\tau})q^{mk^2}\nonumber\\
                        &=& \sum_{k\in\BZ+\frac{n}{2m}}
                            {\rm erf}\left(\sqrt{\frac{-m\tau}{4\pi i}}k
                            \right)q^{mk^2}\,.
\end{eqnarray}
Here, ${\rm erf}(x)$ denotes the usual Gauss error function up to
normalization. To derive the last equality, one has to use the scaling
invariance of the integral measure $\frac{dp}{p}$. Although the set of
functions $\Xi_{n,m}$ and $\widetilde{\Xi}_{n,m}$ closes under the
$S$-transformation, they do not form a representation of the full
modular group, since the $\widetilde{\Xi}_{n,m}$ do not close under $T$.
This means that they do not have a good power series expansion in $q$
with integer coefficients and powers which differ by integers only. From
this follows that the modular group forms an infinite dimensional
representation by repeated action of $T$ on
$\widetilde{\Xi}_{n,m}$. Therefore we conclude that the
${\cal W}(2,2p-1)$-algebras do not yield RCFTs.
  \par
Similar to the case of the elliptic functions $\Theta_{\lambda,k}$, one may
introduce additional variables which correspond to additional quantum
numbers. For example we could write
\begin{equation}
  \Xi_{n,m}(\tau,z) = \sum_{k\in\BZ+\frac{n}{2m}}\sigma(k,z)q^{mk^2}\,.
\end{equation}
The variable $z$ could belong to the eigenvalue of the additional element
$W_0$ of the Cartan subalgebra, actually to its square, since only the
latter can be determined. From the transformation behavior of the
{\euf su}(2)-$\Theta$-functions \cite{KaPe84} we get
\begin{eqnarray}
  \Xi_{n,m}(\tau+1,z) &=& e^{\frac{\pi in^2}{2m}}\Xi_{n,m}(\tau,z)\,,\\
  \Xi_{n,m}(-\frac{1}{\tau},z\tau^2-\frac{\tau}{2m}) &=&
    \sqrt{\frac{-i\tau}{2m}}\sum_{n'\,{\rm mod}\,2m}\sin(-2\pi\frac{nn'}{2m})
    \Xi_{n',m}(\tau,z)\,.
\end{eqnarray}
Indeed, this set of functions forms a finite dimensional representation
of the modular group. But the presence of an additional quantum number
indicates that the chiral symmetry algebra is not yet maximally extended.
Some remarks on this may be found in \cite{FHW93,FrWe93}.
  \par
  \subsection{Characters of the triplet algebras {\myboldmath
  $\w(2,2p-1,2p-1,2p-1)$}}
  \pn
We now study the triplet algebras. Again, we construct $\w$-characters
by summing up the Virasoro characters of degenerate representations
whose highest weights differ by integers. In addition, we have to take
care of multiplicities coming from the {\euf su}$(2)$ symmetry.
Using the isomorphism between fields and Fourier modes which span the
Hilbert space of the vacuum representation, one easily sees that the
multiplicity of the Virasoro HWR on $\vac{h_{2k+1,1}}$ is $2k+1$.
In particular, the multiplicity for $h_{3,1} = 2p-1$, the dimension of
the additional primary fields, is 3 as it should be.
The Virasoro characters are due to Feigin and Fuks \cite{FeFu83}
\begin{equation}
  \chi^{{\em Vir}}_{2k+1,1} = \frac{1}{\eta(q)}\left(q^{h_{2k+1,1}} -
  q^{h_{2k+1,-1}}\right)\,,
\end{equation}
since there is precisely one singular vector in these representations.
The vacuum representation of the $\w$-algebra is then the Hilbert space
\begin{equation}
  {\EH}^{\w}_{\vac{0}} = \bigoplus_{k\in\BZ_+} (2k+1){\EH}^{{\em
  Vir}}_{\vac{h_{2k+1,1}}}\,.
\end{equation}
Therefore, the vacuum character is
\begin{eqnarray}
  \chi^{\w}_{0} &=& \sum_{k\in\BZ_+}(2k+1)\chi^{{\em Vir}}_{2k+1,1}\nonumber\\
                &=& \frac{q^{(1-c)/24}}{\eta(q)}\left(
                    \sum_{k\geq 0}(2k+1)q^{h_{2k+1,1}} -
                    \sum_{k\geq 0}(2k+1)q^{h_{-(2k+1),1}}\right)\nonumber\\
                &=& \frac{q^{(1-c)/24}}{\eta(q)}\left(
                    \sum_{k\geq 0}(2k+1)q^{h_{2k-1,1}} +
                    \sum_{k\geq 1}(-2k+1)q^{h_{-2k+1,1}}\right)\\
                &=& \frac{q^{(1-p)^2/4p}}{\eta(q)}
                    \sum_{k\in\BZ} (2k+1)q^{(1 - (2k+1)p)^2 -
                    (1-p)^2)/4p}\nonumber\\
                &=& \frac{1}{\eta(q)}
                    \sum_{k\in\BZ}(2k+1)q^{(2pk+(p-1))^2/4p}\nonumber\,.
\end{eqnarray}
This can be expressed in terms of $\Theta$-functions and affine
$\Theta$-functions as
\begin{equation}
  \chi^{\w}_{0} = \frac{1}{p\eta(\tau)}\left((\partial\Theta)_{p-1,p}(\tau)
                  + \Theta_{p-1,p}(\tau)\right)\,.
\end{equation}
But now we are in trouble here, since only the functions
$\Lambda_{\lambda,k} = \Theta_{\lambda,k}/\eta$ are modular forms of
weight zero, while the terms
$(\partial\Lambda)_{\lambda,k} = (\partial\Theta)_{\lambda,k}/\eta$ have
the modular weight 1.
  \par
Let us consider the modular transformation behavior of
$(\partial\Lambda)_{\lambda,k}$ under $S$ and $T$. From
(\ref{eq:theta}) we get the relations
\begin{eqnarray}
  (\partial\Lambda)_{\lambda,k}(\tau + 1) & = &
  \exp\left(2\pi i\left(\frac{\lambda^2}{4k} - \frac{1}{24}\right)\right)
  (\partial\Lambda)_{\lambda,k}\,,\\
  (\partial\Lambda)_{\lambda,k}(-\frac{1}{\tau}) & = &
  (-i\tau)\sqrt{\frac{2}{k}}\sum_{1\leq\lambda'\leq k-1}
  \sin\left(\frac{\pi\lambda\lambda'}{k}\right)
  (\partial\Lambda)_{\lambda',k}\,.
\end{eqnarray}
Note the occurrence of a term $\tau$, which cannot be written as a power
series in $q$. We define
$(\nabla\Lambda)_{\lambda,k} \equiv -\tau(\partial\Lambda)_{\lambda,k}$,
which have the modular properties
  \begin{eqnarray}
  (\nabla\Lambda)_{\lambda,k}(\tau + 1) & = &
  \exp\left(2\pi i\left(\frac{\lambda^2}{4k} - \frac{1}{24}\right)\right)
  \left((\nabla\Lambda)_{\lambda,k}-(\partial\Lambda)_{\lambda,k}\right)\,,\\
  (\nabla\Lambda)_{\lambda,k}(-\frac{1}{\tau}) & = &
  -i\sqrt{\frac{2}{k}}\sum_{1\leq\lambda'\leq k-1}
  \sin\left(\frac{\pi\lambda\lambda'}{k}\right)
  (\partial\Lambda)_{\lambda',k}\,.
\end{eqnarray}
It is remarkable, that the $T$-transformation is no longer diagonal.
In some cases the $h$-values of the allowed HWRs are
explicitly known. These are
$\w(2,3,3,3)$ at $c = -2$, with the only possible highest weights
$h\in\{-1/8,0,3/8,1\}$, and $\w(2,5,5,5)$ at $c = -7$, which has HWRs
for $h\in\{-1/3,-1/4,0,5/12,1,7/4\}$ only.
With these data one can solve the modular differential
equation to find the characters. The result is up to base changes the
same.
  \par
We would like to note that one can formally read off the possible
representations from the conformal grid of minimal models in the
following way: The possible $h$-values of a minimal model with
$c = c_{p,q}$ are given by $h_{r,s} = \frac{(pr-qs)^2-(p-q)^2}{4pq}$
with $1\leq r<q$ and $1\leq s<p$. One obtains the $h$-values for a
$c_{p,1}$-model including all inequivalent representations to the same
highest weight from the conformal grid of $c_{3p,3}$.
  \par
For simplicity, we concentrate now on the case $c=-2$, i.e.\ $p=2$.
We first assume the usual form of the characters,
\begin{equation}
   \chi_i = q^{h_i-c/24}\sum_{l=0}^{\infty} b_{i,l}q^l\,,
\end{equation}
where $h_i$ is given by $h_{1,i} = \frac{i^2-2ip+2p-1}{4p}$.
Solving the modular differential equation yields up to multiplicative
prefactors the characters
\begin{equation}\begin{array}{rcl}
  \chi_1 & = & A\Lambda_{1,2} + B(\partial\Lambda)_{1,2}\,,\\
  \chi_2 & = & \Lambda_{0,2}\,,\\
  \chi_3 & = & A'\Lambda_{1,2} + B'(\partial\Lambda)_{1,2}\,,\\
  \chi_4 & = & \Lambda_{2,2}\,,\\
  \chi_5 & = & \frac{1}{2}\Lambda_{1,2} - \frac{1}{2}(\partial\Lambda)_{1,2}\,.
\end{array}\end{equation}
Therefore, $\chi_1$, $\chi_3$ and $\chi_5$ are linear dependent.
If $\chi_1$ is supposed to belong to the vacuum representation, its
coefficient to $q$ must vanish, i.e.\ $b_{1,1} = 0$. This forces
$A = B = 1/2$, if one also requires $b_{1,0} = 1$.
  \par
We now need one further, linear independent solution. We make the ansatz
\begin{equation}
  \tilde{\chi}_3 = \log(q)q^{1/12}\sum_{l=0}^{\infty}\tilde{b}_{3,l}q^l\,.
\end{equation}
Inserting this into the modular differential equation, we get
\begin{equation}
  \tilde{\chi}_3 = (\nabla\Lambda)_{1,2}\,,
\end{equation}
where from now on we define the characters as functions in $q$, i.e.\
$(\nabla\Lambda)_{\lambda,k} \equiv -\frac{\log(q)}{2\pi i}
(\partial\Lambda)_{\lambda,k}$.
Indeed, our result is exactly the same as what we got from the explicit
calculation of the vacuum character and its $S$-transformation.
Replacing $\chi_3$ by $\tilde{\chi}_3$, we obtain the $S$-matrix:
\begin{equation}\label{eq:smatl}
  S = \left(\begin{array}{ccccc}
      0 & \frac{1}{4} &  \frac{i}{2} & -\frac{1}{4} &  0 \\
      1 & \frac{1}{2} &            0 &  \frac{1}{2} &  1 \\
     -i &           0 &            0 &            0 &  i \\
     -1 & \frac{1}{2} &            0 &  \frac{1}{2} & -1 \\
      0 & \frac{1}{4} & -\frac{i}{2} & -\frac{1}{4} &  0
  \end{array}\right)\,.
\end{equation}
This matrix has, as it should be, det$(S) = 1$, but is neither
symmetric nor real nor unitary. Nonetheless is satisfies $S^2 = \Bid$.
Most disturbing, it has no row/column with only non vanishing entries.
We collect our intermediate results.
  \medskip\pano
  {\sc Proposition 2.} {\em Let $p\in\BN$. Then there exists at
  $c = c_{p,1} = 13 - 6(p + p^{-1})$ a $\w(2,2p-1,2p-1,2p-1)$.
  There are precisely $3p-1$ HWRs with highest weights
  $h_{1,s}, 1\leq s\leq 3p-1$. Of them $2\times(p-1)$ HWRs have
  pairwise identical highest weights, further $p-1$ highest weights
  differ from these pairs by positive integers which are the levels of
  the singular vectors. A basis for the characters is given by
  ($\eta^{-1}$ times) the functions $\{\Theta_{\lambda,p},
  (\partial\Theta)_{\mu,p},(\nabla\Theta)_{\mu,p} |
  0\leq\lambda,\mu\leq(2p-1), \mu\neq 0,p \}$.}
  \medskip\par
We already noted that the functions
$(\nabla\Theta)_{\mu,p}$ lead to a non diagonal
$T$-matrix. It decomposes into blocks similar to Jordan cells, but
which also mix characters whose corresponding highest weights differ by
integers. For our example the $T$-matrix is (in the same basis)
\begin{equation}
  T = \left(\begin{array}{ccccc}
 \exp(\pi i/6)&               &             &                &              \\
              &\exp(-\pi i/12)&             &                &              \\
-\exp(\pi i/6)&              0&\exp(\pi i/6)&               0&\exp(\pi i/6) \\
              &               &             &\exp(11\pi i/12)&              \\
              &               &             &                &\exp(13\pi i/6)
  \end{array}\right)\,.
\end{equation}
Nonetheless, this matrix satisfies together with the $S$-matrix
(\ref{eq:smatl}) the relation $(ST)^3 = \Bid$. This is very important in
order to have modular invariance of the CFT. It is easy to see that the
statement is true for the general case of proposition 2.
  \par
But what are the ``physical'' characters? This question will be answered
by enforcing modular invariance of the partition function. From
our discussion of the modular properties of the characters we know
that the following expression is modular invariant:
\begin{equation}\label{eq:zlog}
  Z_{{\em log}}[p] = \alpha\sum_{\lambda=0}^{2p-1}|\Theta_{\lambda,p}|^2
                   + \beta\sum_{{\mu=1\atop\mu\neq p}}^{2p-1}\frac{1}{2}\left(
                     (\partial\Theta)_{\mu,p}(\nabla\Theta)^*_{\mu,p} +
                     (\nabla\Theta)_{\mu,p}(\partial\Theta)^*_{\mu,p}
                     \right)\,,
\end{equation}
where $\alpha,\beta$ are yet free constants. The partition function will
only be physical relevant, i.e.\ with integer coefficients only, if
$\alpha,\beta\in\BZ/2$.
  \par
According to proposition 2 we introduce
general linear combinations for the characters of each triplet of
degenerate HWRs $(h,h,h+k)$. There is only one such triplet, $(0,0,1)$,
in our example of $p=2$. Then the ansatz is
$$
  \begin{array}{lcl}
  \chi_0 & = & \Theta_{0,2}/\eta\,,\\
  \chi_2 & = & \Theta_{2,2}/\eta\,,\\
  \chi_1^0 & = & (a_0\Theta_{1,2} + b_0(\partial\Theta)_{1,2} +
                 c_0(\nabla\Theta)_{1,2})/\eta\,,\\
  \chi_1^+ & = & (a_+\Theta_{1,2} + b_+(\partial\Theta)_{1,2} +
                 c_+(\nabla\Theta)_{1,2})/\eta\,,\\
  \chi_1^- & = & (a_-\Theta_{1,2} + b_-(\partial\Theta)_{1,2} +

                 c_-(\nabla\Theta)_{1,2})/\eta\,.
  \end{array}
$$
One sees immediately that $a_i,b_i$ have to be real and $c_i$ imaginary
to obtain a (possibly) physical relevant partition function.
Putting $a_0\neq 0$ gives a contradiction. With $a_0 = 0$
one finds the solution $a_+ = a_- = 1$,
$b_-=-b_+$ and $c_+ = -c_- = \pm ib_+$. Then also
$\pm c_0 = -\sqrt{2}ib_+$ and $\pm b_0 = \sqrt{2}b_+$ are fixed up to a
common sign. We see that logarithmic terms occur in {\em all\/} HWRs of
the triplet, since otherwise we cannot cancel terms not allowed in
(\ref{eq:zlog}). Moreover, the $S$-matrix now gets a block structure,
since one $\w$-family, the one with the unphysical multiplicity $\sqrt{2}$,
decouples from the remaining ones. It is remarkable that the four
remaining HWRs obey a good fusion algebra under themselves, if one
applies the Verlinde formula to the corresponding block of the
$S$-matrix, which is (including the free signs)
\begin{equation}
  S =
  \left(\begin{array}{ccccc}
    \mp 1&      0&      0&      0&      0\\
        0&    1/2&    1/2&    1/2&    1/2\\
        0&    1/2&\pm 1/2&   -1/2&\mp 1/2\\
        0&    1/2&   -1/2&    1/2&   -1/2\\
        0&    1/2&\mp 1/2&   -1/2&\pm 1/2
  \end{array}\right)\,.
\end{equation}
Here, the third row/column corresponds to the vacuum representation with
$\chi_1^+$ and the character vector is
$(\chi_1^0,\chi_0,\chi_1^+,\chi_2,\chi_1^-)^t$. This new $S$-matrix is
symmetric and unitary. The remaining constants are fixed by the $T$-matrix.
It follows $b_+=\pm 1$, where a sign change corresponds to an exchange
of the third and fifth row/column. Choosing $b_+ = 1$ selects $\chi_1^+$
as vacuum character, whose  non logarithmic part is -- up to an overall
multiplicity of 2 -- identical to our first result.
The $T$-matrix then reads
\def\eep#1{\exp(\pi i#1)}
\def\eem#1{\exp(-\pi i#1)}
\def\iv{\frac{i}{4}}
\def\ih{\frac{i}{2}}
\def\is{\frac{i}{2\sqrt{2}}}
{\small
\begin{equation}
  T =
  \left(\begin{array}{ccccc}
    (1-\iv)\eep{/6}&          0&    \iv\eep{/6}&   -\is\eep{/6}&        0\\
                  0&\eep{11/12}&              0&              0&        0\\
        \iv\eep{/6}&          0&(1-\iv)\eep{/6}&    \is\eep{/6}&        0\\
        \is\eep{/6}&          0&   -\is\eep{/6}&(1+\ih)\eep{/6}&        0\\
                  0&          0&          0&                  0&\eem{/12}
  \end{array}\right)\,.
\end{equation}
}
  \par
The enlarged multiplicity cannot be avoided, since otherwise the
partition function (\ref{eq:zlog}) would have rational non integer
coefficients. In the general case, the enlarged multiplicity is
just $p$, and stems from the degenerate solutions of chiral vertex
operators of same conformal dimension (as explained in section 1).
In general we now have
  \medskip\pano
  {\sc Proposition 3.} {\em Let $p\in\BN, p\geq 2$. Then there are
  the following HWRs and characters for the $\w(2,2p-1,2p-1,2p-1)$-algebra:
  A unique HWR to the highest weight of minimal energy,
  $h_{{\it min}} = h_{1,p} = h_{0,0} = -(p-1)^2/4p$, with
  character $\chi_p = \Theta_{0,p}/\eta$, one HWR to
  $h_{1,2p} = h_{1,0}$ with character $\chi_0 = \Theta_{p,p}/\eta$, both
  do not contain any null states.
  Moreover, triplets of HWRs to $(h_{1,s}=h_{1,2p-s},h_{1,2p+s})$ with
  inequivalent \footnote{
  Note, that the singular vector of the HWR to $h_{1,s}$ has level $h_{1,-s} =
  h_{1,2p+s}$, but the singular vector of the HWR to $h_{1,2p-s}$ has
  level $h_{1,s-2p} = h_{1,4p-s}$.}
  characters $(\chi_s^+,\chi_s^0,\chi_s^-)$ where
  $\chi_s^{\pm} = (\Theta_{p-s,p}\pm(\partial\Theta)_{p-s,p}\pm
  i(\nabla\Theta)_{p-s,p})/\eta$ and $\chi_s^0 =
  \sqrt{2}((\partial\Theta)_{p-s,p}-i(\nabla\Theta)_{p-s,p})/\eta$.
  The ``diagonal'' partition function is then}
  \begin{equation}
    Z_{{\it log}}[p] = \frac{1}{\eta\bar{\eta}}\left(|\chi_0|^2 + |\chi_p|^2
    +\sum_{1\leq s\leq p-1}\left(|\chi_s^0|^2 + \chi_s^+(\chi_s^-)^*
    +\chi_s^-(\chi_s^+)^*\right)\right)\,.
  \end{equation}
  \medskip\par
Noting that $(\partial\Theta)_{0,p} = (\nabla\Theta)_{p,p} \equiv 0$,
one can write all characters in the same form.
The resulting $S$-matrix has block structure. $p-1$
logarithmic representations with characters $\chi_s^0,1\leq s\leq p-1$
decouple from the remaining $2(p-1)+2 = 2p$ ``regular'' representations.
For the latter we already derived abstract fusion rules in section 1.
We remark that the quantum dimensions of the decoupling representations
all vanish. This and the block structure of the $S$-matrix show that the
Verlinde formula is no longer valid. We only can calculate the fusion
rules for the $2(p-1)$ regular representations, if we use the
corresponding block of the $S$-matrix. Nonetheless, the fusion rules
calculated from the ``regular'' block of the $S$-matrix are well behaved.
In the last section we will discuss their physical relevance. For our
example we obtain from the Verlinde formula, applied to the $4\times4$
block,
  \begin{equation}\label{eq:Vfus}
  \begin{array}{rclcl}
{}[-\frac{1}{8}]&\times&[-\frac{1}{8}] & = & [0]        \,,\\
{}[-\frac{1}{8}]&\times&[0]        & = & [-\frac{1}{8}] \,,\\
{}[-\frac{1}{8}]&\times&[\frac{3}{8}]  & = & [1]        \,,\\
{}[-\frac{1}{8}]&\times&[1]        & = & [\frac{3}{8}]  \,,\\
{}       [0]&\times&[0]        & = & [0]        \,,
  \end{array}
  \begin{array}{rclcl}
{}       [0]&\times&[\frac{3}{8}]  & = & [\frac{3}{8}]  \,,\\
{}       [0]&\times&[1]        & = & [1]        \,,\\
{} [\frac{3}{8}]&\times&[\frac{3}{8}]  & = & [0]        \,,\\
{} [\frac{3}{8}]&\times&[1]        & = & [-\frac{1}{8}] \,,\\
{}       [1]&\times&[1]        & = & [0]        \,,\\ {\rm and\ }
{}  [\tilde0]&\times&[\tilde0] & = & [\tilde0]  \,.
  \end{array}
  \end{equation}
The second representation with $h=0$, $[\tilde0]$, completely decouples from
the other representations. This is similar to a well known
phenomenon in $q$-algebras, where representations with vanishing
quantum dimensions are invisible for the other representations. This
also explains the unphysical multiplicity $\sqrt{2}$ for characters
of such representations. Actually, the underlying $q$-algebra structure
admits to representations whose quantum dimensions add up to zero. Thus,
the corresponding  $\w$-algebra representations are
degenerated. For example, we have two representations
$[\tilde0_{\pm}]$ with characters $\chi_{1,0}^+=-\chi_{1,0}^-
=\chi_{1,0}/\sqrt{2}$.
  \par
The case $p=1$ is trivial, there are no logarithmic representations. It is
just $\widehat{{\euf su}(2)}$, the simplest non-abelian
infinite dimensional Lie algebra $A_1^{(1)}$, with $c=1$. In particular,
$Z_{{\log}}[1] = Z[1]$, where $Z[x]$ denotes the standard Gaussian U(1)
partition function for a free field compactified with radius
$R=\sqrt{(x/2)}$, usually denoted $Z(R)$. This means that our logarithmic
CFT reduces to the Gauss
model at the multi-critical point of radius $1/\sqrt{2}$.
But R.~Dijkgraaf and E.~\&~H.~Verlinde \cite{DVV88} have proven that
there are {\em no\/} marginal deformations, which can lead out of the
known moduli space of $c=1$ CFTs. There is one field of marginal
dimension, $\phi_{2,p-1}$ with $h_{2,p-1} = 1$, which belongs to the
(extended) conformal grid of section 1. Since the first label is even,
it has vanishing self coupling, which is necessary for a marginal
operator to be integrable. But this field does not exist for $p=1$,
since all fields $\phi_{r,s}$
with $r=0$ or $s=0$ decouple completely from the physical Hilbert space
due to annihilation by the BRST operator. Thus, we indeed cannot go from
the moduli space of regular $c=1$ CFTs to the logarithmic CFTs with
$c_{{\em eff}}=1$ via marginal deformations. If we finally note that the
partition function of proposition 3 also allows non diagonal
decompositions, we have
  \medskip\pano
  {\sc Proposition 4.} {\em The moduli space of logarithmic CFTs with
  $c_{{\it eff}} = 1$ is generic one dimensional and not connected to
  the moduli space of regular $c=1$ CFTs.
  The partition function of a logarithmic CFT is for $(p,q)=1$ given by
  \begin{equation}
    Z_{{\it log}}[p/q] = \frac{1}{\eta\bar{\eta}}\left(|\chi_0|^2
                       + |\chi_{pq}|^2
                       + \sum_{1\leq s\leq pq-1}\left(\chi_s^0(\chi_{s'}^0)^*
                       + \chi_s^+(\chi_{s'}^-)^*
                       + \chi_s^-(\chi_{s'}^+)^*\right)\right)\,,
  \end{equation}
  where $s=pn-qm$ mod $2pq$ implies $s'=pn+qm$ mod $2pq$.}
  \medskip\par
The connected part of the moduli space of $c=1$ theories has an exact
copy of logarithmic theories in the following manner: First, one writes
\begin{equation}
  Z_{{\em log}}[x] = \left(1 + \frac{2x^2}{\pi i}\frac{\partial}{\partial x}
  \right)Z[x]\,,
\end{equation}
which by the way defines $Z_{{\em log}}[x]$ for arbitrary, not
necessarily rational $x$. In the same way we obtain the partition
function of the $\BZ_2$-orbifolds of the logarithmic theories by
applying $(1+\frac{2x^2}{\pi i}\partial_x)$ to $Z_{{\em orb}}[x]$,
\begin{equation}
  Z_{{\em log,orb}}[x] = \left[
  \left(1+\frac{2x^2}{\pi i}\frac{\partial}{\partial x}\right)Z[x]
  %+ 2Z[4] - Z[1])/2\,.
  + \left.\left(1+\frac{2y^2}{\pi i}\frac{\partial}{\partial y}
  \right)Z[y]\right|_{y=4} - Z[1]\right]/2\,.
\end{equation}
The corresponding $\w$-algebras, which exist at points of enhanced
symmetry analogous to the regular case, are the following:
To $Z_{{\em log}}[p], p\in\BN$ belongs a $\w(2,(2p-1)^{\otimes 3})$, whose
$\BZ_2$-orbifold contains a $\w(2,6p-2)$, the $\BZ_2$-orbifold of
$\w(2,2p-1)$ where the singlet field is given by $W = W_0 + W_+ + W_-$ and
the orbifold is obtained by identifying $W$ with $-W$. Since the structure
constant $C_{WW}^W$ does not vanish for the triplet, the $\BZ_2$-orbifold
of the triplet should be given by the identifications
$W_0\leftrightarrow -W_0$, $W_+\leftrightarrow -W_-$, and
$W_-\leftrightarrow -W_+$ such that one field, e.g.\ $\tilde W =
W_+ - W_-$ survives. The orbifold would then be a $\w(2,2p-1,6p-2)$.
If $p$ is a complete square,
$p = n^2$, these algebras can be extended by a field of dimension
$h_{2n+1,1} = p(n^2+n)-n = n^4 + n^3 - n$.
%This must remain true
%for $p\neq n^2$, since $\w(2,6p-2)$ is not maximally extended in the way
%that it yields a (logarithmic) RCFT \cite{EHH93b}.
In the same manner
one can write down the logarithmic analogs of the three exceptional
$c=1$ partition functions. Setting $D_x=\frac{2x^2}{\pi i}\partial_x$,
the exceptional logarithmic partition functions simply read
\begin{eqnarray}
  Z_{{\em log},E_6} &=& \frac{1}{2}\left(\sum_{x\in\{4,9,9\}}(1+D_x)Z[x]
    -Z[1]\right)\,,\\
  Z_{{\em log},E_7} &=& \frac{1}{2}\left(\sum_{x\in\{4,9,16\}}(1+D_x)Z[x]
    -Z[1]\right)\,,\\
  Z_{{\em log},E_8} &=& \frac{1}{2}\left(\sum_{x\in\{4,9,25\}}(1+D_x)Z[x]
    -Z[1]\right)\,.
\end{eqnarray}
In this way, the full $c=1$ moduli space is
recovered in the ``logarithmic'' regime. There are no other linear
combinations possible, since the non-logarithmic part of the partition
function has to satisfy the usual requirements to be physical relevant,
which only yield the known $c=1$ solutions.
  \par
Of course, there could be higher powers of logarithmic terms.
All expressions of the form $(\sum_{n\in\BZ_+}a_n D^n_x)Z[x]$ are
modular invariant. Fortunately, as mentioned above, this presumably cannot
happen for theories with $c_{{\em eff}} \leq 1$ (see also \cite{PhD}).
  \par
We conclude with a remark on $N=1$ supersymmetric theories. The explicit
known examples \cite{BEHH92,EHH93b} as well as the general results on
the modular properties of characters make it clear that $N=1$ CFTs will
have the same structure. One finds again logarithmic theories
(with $c_{{\em eff}} = 3/2$), which have a completely analogous
representation theory. This analogy extends the similarity of the
representation
theory of the already known $N=0,1$ RCFTs \cite{Flo93}.
But as already observed in other cases, such results do not extend to
$N=2$, since there no rational like structure can be found
(for examples see \cite{Blu93}). It remains the conjecture that for
$N=2$ rationality of a CFT implies its unitarity.
%
%%< DISCUSSION: POLYMERS >%%%%%%%%%%%%%%%%%%%%%%%%%%%%%%%%%%%%%%%%%%%%%%%%
%
  \par
  \mysection{Discussion: Polymers}
  \pn
Finally, we would like to discuss the consequences of our results in
relation to the theoretical understanding of two dimensional polymers.
In an early work \cite{DuSa87}, B.~Duplantier and H.~Saleur derived that
the full partition function of two dimensional polymers in the dense
phase is trivially zero. But they also showed that the partition
function of a single self-avoiding polygon (a dense loop polymer)
homotopic to zero on a torus is given by
  \begin{equation}
    \tilde{X} = -\frac{1}{4\pi}\eta^2(q)\eta^2(\bar{q})\log(q\bar{q})
    = \left.\left(D_xZ[x]\right)\right|_{x=1/2}\,,
  \end{equation}
which is the modular invariant part of of the summed number of configurations.
Surprisingly, this is precisely the logarithmic part of our $c=-2$ partition
function.
  \par
In a later work \cite{Sal92} H.~Saleur studied the polymer problem again
and found a hidden $N=2$ supersymmetry by modeling it via a $\eta$-$\xi$
system. This model has different sectors which combine into two modular
invariant partition functions, one for even number of non-contractable
polymers, which is $Z_{{\em even}} = \frac{1}{2}Z[2]$, and one for odd number
of non-contractable polymers, which is $Z_{{\em odd}} = Z[8] -
\frac{1}{2}Z[2]$.
The latter corresponds to a so called twisted sector of the $\eta$-$\xi$ system
with $\BZ_4$ symmetry. We find now the following:
$Z_{{\em even}}$ equals the regular part of our partition function of
the $c=-2$ model, $Z_{{\em odd}}$ is identical to the partition function of
the $c=-2$ model with twisted bosons, i.e.\ the bosonic fields have now
half-integer Fourier modes. One obtains the partition function from the
modular properties of the Jacobi-Riemann elliptic functions
$\Theta_{\lambda,p}$ with half-integer index $\lambda$. In fact, the
twisted $\w(2,3,3,3)$-model allows only two HWRs which have $h=-5/32$ and
$h=3/32$. They have the characters $\chi_{1/2} = \Theta_{\frac{1}{2},2}/\eta$
and $\chi_{3/2} = \Theta_{\frac{3}{2},2}/\eta$ respective. The resulting
partition function is just $Z_{{\em odd}}$.
  \par
As a result, the modular invariant partition function of our
$\w(2,3,3,3)$-model at $c=-2$ naturally contains {\em all} different
partition functions of two-dimensional dense polymers, {\em including}
the logarithmic part which comes from polymers homotopic to zero. The
hidden $N=2$ supersymmetry in form of a $\eta$-$\xi$ system still is visible
in the behavior of the screening currents: $Q$ and the current
$\tilde{J}$ corresponding to $\tilde{Q}$ create states similar but not
equal to a usual $\eta$-$\xi$ system.
  \par
The difference between our approach and the one with $N=2$ supersymmetry
is that the latter does not detect the states corresponding to the
logarithmic operators. But as shown in \cite{Gur93}, these operators
cannot be avoided, if the OPE and the construction of conformal blocks
should remain consistent. Without these additional states one ends up with
a partition function which is equal to an ordinary $c=1$ Gaussian model.
Only the $h$-values change due to the shift of the central charge to $c=-2$.
Certainly, the structure of the CFTs at $c=1$ and $c=-2$ is different. This
is not visible in the supersymmetric approach, presumably because the
corresponding chiral algebra is not maximal. The additional states are
naturally accounted for in our model. Surprisingly this just leads to an
additional term in the partition function which represents the polymers
homotopic to zero.
  \par
One possible way to find the missing states in the $\eta$-$\xi$ system is
the observation that the naive $L_0$ from $T(z) = :\eta\partial\xi:$ is
diagonal. To recover the correct Jordan cell structure, which only is present
in the Neveu-Schwarz sector, one can modify $L_0$ in the way
\be\label{eq:newL0}
  L_0 = \sum_{m=1}^{\infty} m(\xi_{-m}\eta_{m} + \eta_{-m}\xi_{m}) + \eta_0\,,
\ee
which does not change any commutators, since $\{\xi_m,\eta_n\} =
\delta_{m+n,0}$ and thus $[L_m,\eta_n] = n\eta_{m+n}$.
By this, we correctly obtain $L_0(\xi_0\vac{0}) = \vac{0}$, such that the
two states $\vac{0;0}=\vac{0}$ and $\vac{0;1}=\xi_0\vac{0}\neq 0$ span the
Jordan cell. This also justifies a posteriori our definition of characters for
non-diagonal $L_0$, as given in the second section.
Actually, the form given above naturally matches the definition of $L_0$
as matrix, given in section 2,
where the non-diagonal part only contributes to the states on $\vac{0;1}$.
But now, (\ref{eq:newL0}) gives $L_0$ in such a way that taking traces is
still well defined, and thus characters can be defined as usual.
This correction of $L_0$ is very reminiscent
of a similar ``explicit'' visibility of the related quantum group structure in
the CFT of Liouville theory \cite{GoSi91}, where also $L_0$ is non-diagonal.
  \par
Let us finally discuss the operator algebra. In \cite{Sal92}, the operators
of the $c=-2$ CFT have been identified with the $L$-leg operators in the
following way: In the antiperiodic (Ramond) sector, we have fields of
conformal weights $h_{1,2+2l}$ which correspond to $4l$-leg operators.
Their fermion number is $F\equiv l$ mod 2. Our model collects these
operators into two $\w$-conformal families, $[-\frac{1}{8}]$ with $F=0$, and
$[\frac{3}{8}]$ with $F=1$. Due to the fermion number, the first HWR is a
singlet, the second a doublet representation. The periodic (Neveu-Schwarz)
sector contains fields of conformal weights $h_{1,3+4k}$ and $h_{1,5+4k}$
which correspond to the $(4l+2)$-leg operators for $l=2k$ and $l=2k+1$
respectively. Again, our model collects these operators into two families,
$[0]$ for $l$ even, and $[1]$ for $l$ odd. For the sake of completeness,
we just mention that the so called $\BZ_4$ sector in \cite{Sal92}
coincides with the twisted sector of our model, which splits into the
two families $[-\frac{3}{32}]$ and $[\frac{5}{32}]$ which correspond to
the $(4l+1)$-leg operators and the $(4l+3)$-leg operators, where the
first family again collects operators for $l$ even, the second collects
operators for $l$ odd. In all families, the fields are descendants of the
one with the lowest value $L$ of legs. Thus, we obtain the following splitting
of the operators modulo 8:
  \begin{eqnarray}
    {\rm Ramond:}        & L\equiv 0(8)\,,  &
                           \Phi_L\in [{\ts-\frac{1}{8}}]\,,\nonumber\\
                         & L\equiv 4(8)\,,  &
                           \Phi_L\in [{\ts\frac{3}{8}}] \,,\nonumber\\
    {\rm Neveu-Schwarz:} & L\equiv 2(8)\,,  &
                           \Phi_L\in [0]       \,,\\
                         & L\equiv 6(8)\,,  &
                           \Phi_L\in [1]       \,,\nonumber\\
    \BZ_4 {\rm \ sector:}& L\equiv 1,3(8)\,,&
                           \Phi_L\in [{\ts-\frac{3}{32}}]\,,\nonumber\\
                         & L\equiv 5,7(8)\,,&
                           \Phi_L\in [{\ts\frac{5}{32}}]\,.\nonumber
  \end{eqnarray}
Due to the more complicated modular transformation behavior of the
characters in the $\BZ_4$ sector, which involves alternating
$\Theta$-functions $\tilde{\Theta}_{\lambda,k}=
\sum_{n\in\BZ}(-)^nq^{(4kn+\lambda)^2/4k}$, and therefore a second set
of characters $\tilde{\chi}(\tau) \equiv \chi(\tau + 1)$, the splitting
modulo 8 can be made rigorous by considering $(\chi\pm\tilde{\chi})/2$.
  \par
The fusion rules (\ref{eq:Vfus}) respect the sector structure derived
by H.~Saleur \cite{Sal92}. The $4l$-leg operators build the partition
function which (on the lattice put on a torus) sums over dense coverings
by an even number of non-contractable polymers that cross both periods
in the total
an odd number of times. Of particular interest are the representations
for the $(4l+2)$-leg operators. Their characters combine to the part of the
partition function which (on the lattice put on a torus) sums over dense
coverings by an even number of non-contractable polymers that cross one
period an odd number of times. But they also yield the part of the partition
function which is the number of configurations of a single contractable densely
covering polymer. Actually, the naive partition function for doubly periodic
boundary conditions vanishes, but its derivative is precisely this logarithmic
part of our full partition function. The twisted sector partition function
(on the lattice put on a torus) sums over dense coverings by an odd number
of non-contractable polymers.
  \par
We see, that the Neveu-Schwarz sector also contains the contribution
of one dense, contractable loop. This contribution stems from the Jordan
cell structure, i.e.\ the fact that application of $L_0$ produces new states.
In the Temperley-Lieb picture of polymer configurations it is natural to
use a single dense contractable polymer as a possible groundstate from which
non-contractable configurations can be created by action of the
Termperley-Lieb operators. In fact, the BRST operator
\be
  Q_{{\rm BRST}} = \oint\frac{dz}{2i\pi}\eta(z)
\ee
is the polymer creating operator, which may as well act on the sector
without polymers as on the sector with exactly one contractable polymer as
groundstates. More than one contractable polymer always yields zero due to the
special property of the corresponding Temperley-Lieb algebra
that $e^2 = \delta e$ with $\delta = 0$. The BRST property $Q_{{\rm BRST}}^2
= 0$ remains valid in the sector with one single contractable polymer.
  \par
Physically, the three HWRs $[0]$, $[1]$, and $[\tilde 0]$ correspond to
three different ways of counting states build on groundstates with zero or one
contractable dense loop. The actual combinations are selected by modular
invariance of the partition function. $[0]$ and $[1]$ count states from both
groundstates with and without a sign for states with contractable loop.
$[\tilde 0]$ may be interpreted as the HWR to the density field.
  \par
Since it is believed that the higher members of the $c_{1,p}$ series
describe multi-critical polymers, we conjecture that the corresponding
$\w(2,(2p-1)^{\otimes 3})$ models are the right candidates for these
physical systems. In fact, the Jordan cell structure of $L_0$ is a common
feature of all members of the $c_{1,p}$ series, which is naturally
incorporated in these multiplet $\w$-algebras.
  \bigskip\pano
{\bf Acknowledgment:}
I would like to thank C.~Gomez, A.~Honecker, W.~Nahm, G.~Sierra
and R.~Varnhagen for useful discussions. In particular I thank
W.~Eholzer, H.G.~Kausch and F.~Rohsiepe
for numerous illuminating discussions and comments.
This work has been supported partly by
the Deutsche Forschungsgemeinschaft and partly by the European
Scientific Network n$^{{\rm o}}$ ERB CHRX CT 920069.
% \newpage
%
%%< REFERENCES >%%%%%%%%%%%%%%%%%%%%%%%%%%%%%%%%%%%%%%%%%%%%%%%%%%%%%%%%%%
%
  

\begin{thebibliography}{BFKNRVl}
  {\footnotesize
\bibitem[BPZ83]{BPZ83} {\sc A.A. Belavin, A.M. Polyakov, A.B. Zamolodchikov},
%  {\em Infinite conformal symmetry in two-dimensional quantum field theory},
   Nucl. Phys. {\bf B241} (1984) 333
\bibitem[Blu93]{Blu93} {\sc R. Blumenhagen},
%  {\em $N=2$ Supersymmetric $\w$-Algebras},
   Nucl. Phys. {\bf B405} (1993) 744,\\
   {\sc R. Blumenhagen, R. H{\"u}bel},
%  {\em A Note on Representations of $N=2$ ${\cal SW}$-Algebras},
%  Universit"at Bonn Preprint BONN-TH-94-08, hep-th/9407068 (1994)
   Mod. Phys. Lett. {\bf A9} (1994) 3193
\bibitem[BEH$^2$92]{BEHH92} {\sc R. Blumenhagen, W. Eholzer, A. Honecker,
   R. H{\"u}bel},
%  {\em New $N=1$ Extended Superconformal Algebras with Two and Three
%  Generators},
   Int. Jour. Mod. Phys. {\bf A7} (1992) 7841
\bibitem[BEH$^3$94]{BEHHH94} {\sc R. Blumenhagen, W. Eholzer, A. Honecker,
   K. Hornfeck, R. H{\"u}bel},
%  {\em Unifying $\w$-Algebras},
   Phys. Lett. {\bf B332} (1994) 51-60,
%  {\em Coset Realization of Unifying $\w$-Algebras},
%  INFN Turin Preprint DFTT-25-94, hep-th/9406203 (1994)
   Int. Jour. Mod. Phys. {\bf A10} (1995) 2367
\bibitem[BFKNRV91]{BFKNRV91} {\sc R. Blumenhagen, M. Flohr, A. Kliem, W. Nahm,
   A. Recknagel, R. Varnhagen},
%  {\em $\w$-Algebras with two and three Generators},
   Nucl. Phys. {\bf B361} (1991) 255
\bibitem[Car86]{Car86} {\sc J.L. Cardy},
%  {\em Operator Content of Two-Dimensional Conformally Invariant Theories}
   Nucl. Phys. {\bf B270} (1986) 186
\bibitem[DV$^2$88]{DVV88} {\sc R. Dijkgraaf, E. Verlinde, H. Verlinde},
%  {\em $c=1$ Conformal Field Theories on Riemann Surfaces},
   Commun. Math. Phys. {\bf 115} (1988) 649
\bibitem[DoFa84]{DoFa84} {\sc V.S. Dotsenko, V.A. Fateev},
%  {\em Conformal Algebra and Multipoint Correlation Functions in 2d
%  Statistical Models},
   Nucl. Phys. {\bf B249}[FS12] (1984) 312,
%  {\em Four-Point Correlation Functions and the Operator Algebra in 2d
%  Conformal Invariant Theories with Central Charge $c\leq 1$},
   Nucl. Phys. {\bf B251}[FS13] (1985) 691,
%  {\em Operator Algebra of Two-Dimensional Conformal Theories with
%  Central Charge $c \leq 1$},
   Phys. Lett. {\bf B154} (1985) 291
\bibitem[DuSa87]{DuSa87} {\sc B. Duplantier, H. Saleur},
%  {\em Exact critical properties of two-dimensional dense selfavoiding walks},
   Nucl. Phys. {\bf B290}[FS20] (1987) 291
\bibitem[EFH$^2$V93]{EFHHV93} {\sc W. Eholzer, M. Flohr, A. Honecker,
   R. H{\"u}bel, R. Varnhagen},
   {\em $\w$-Algebras in Conformal Field Theory},
   to be published in Proc. Trieste Workshop
   {\em Superstrings and Related Topics}, Trieste, July 1993
\bibitem[EH$^2$93]{EHH93b} {\sc W. Eholzer, A. Honecker, R. H{\"u}bel},
%  {\em How complete is the classification of $\w$-symmetries?},
   Phys. Lett. {\bf B308} (1993) 42
\bibitem[EhSk94]{Eho00} {\sc W. Eholzer, N.-P. Skoruppa},
   {\em Modular Invariance and Uniqueness of Conformal Characters},
%  Universit{\"a}t Bonn Preprint BONN-TH-94-16, Max-Planck-Institut f{\"u}r
%  Mathematik Bonn Preprint MPI-94-67, hep-th/9407074 (1994)
   preprint hep-th/9407074, BONN-TH-94-16, MPI/94-67, to appear
   in Commun. Math. Phys.
\bibitem[FeFu82]{FeFu82} {\sc B.L. Feigin, D.B. Fuks},
%  {\em Invariant skew-symmetric differential operators on the line and
%  Verma modules over the Virasoro algebra},
   Funkt. Anal. Appl. {\bf 16} (1982) 114
\bibitem[FeFu83]{FeFu83} {\sc B.L. Feigin, D.B. Fuks},
%  {\em Verma Modules over the Virasoro Algebra},
   Funct. Anal. Appl. {\bf 17} (1983) 241,
%  {\em Verma Modules over the Virasoro Algebra},
   in {\em Topology}, Proc. Leningrad 1982, L.D. Faddeev, A.A. Mal'cev (eds.),
   Lect. Notes Math. {\bf 1060} (1984) 230, Springer Verlag
\bibitem[Fel89]{Fel89} {\sc G. Felder},
%  {\em BRST Approach to Minimal Models},
   Nucl. Phys. {\bf B317} (1989) 215,
   {\em Erratum},
   Nucl. Phys. {\bf B324} (1989) 548
\bibitem[F$^2$K89]{FFK89} {\sc G. Felder, J. Fr{\"o}hlich, G. Keller},
%  {\em Braid Matrices and Structure Constants for Minimal Conformal Models},
   Commun. Math. Phys. {\bf 124} (1989) 647
\bibitem[Flo93]{Flo93} {\sc M. Flohr},
%  {\em $\w$-Algebras, New Rational Models and the Completeness of the
%  $c=1$ Classification},
   Commun. Math. Phys. {\bf 157} (1993) 179
\bibitem[Flo94]{Flo94} {\sc M. Flohr},
%  {\em Curiosities at Effective $c = 1$},
   Mod. Phys. Lett. {\bf A9} (1994) 1071
\bibitem[Flo94PhD]{PhD} {\sc M. Flohr},
   {\em The Rational Conformal Quantum Field Theories in Two Dimensions with
   Effective Central Charge $c_{{\rm eff}} \leq 1$},
   preprint BONN-IR-94-11, (Ph.D. thesis in german, 1994)
\bibitem[FHW93]{FHW93} {\sc M.D. Freeman, K. Hornfeck, P. West},
%  {\em Commuting quantities and exceptional $\w$-algebras},
   Int. J. Mod. Phys. {\bf A8} (1993) 909
\bibitem[FrWe93]{FrWe93} {\sc M.D. Freeman, P. West},
   {\em On the relation between integrability and infinite-dimensional
   algebras},
   preprint hep-th/9303119, KCL-TH-93-1 (1993)
\bibitem[GoSi91]{GoSi91} {\sc C. Goemz, G. Sierra},
%  {\em A Note on Liouville Theory and the Uniformization of Riemann Surfaces},
%  in Proc. Miami 1991, on {\em Quantum Field Theory, Statistical Mechanics,
%  Quantum Groups and Topology}, 115-122,
   Phys. Lett. {\bf B225} (1991) 51, Int. J. Mod. Phys. {\bf A6} (1991)
   2045-2074
\bibitem[Gur93]{Gur93} {\sc V. Gurarie},
%  {\em Logarithmic Operators in Conformal Field Theory},
   Nucl. Phys. {\bf B410} (1993) 535
\bibitem[KaPe84]{KaPe84} {\sc V.G. Kac, D.H. Peterson},
%  {\em Infinite-Dimensional Lie Algebras, Theta Functions and Modular Forms},
   Adv. Math. {\bf 53} (1984) 125
\bibitem[Kau91]{Kau91} {\sc H.G. Kausch},
%  {\em Extended Conformal Algebras Generated by a Multiplet of Primary
%  Fields},
   Phys. Lett. {\bf B259} (1991) 448
\bibitem[Kau9?]{Kau95} {\sc H.G. Kausch},
   {\em Chiral Algebras in Conformal Field Theory},
   preprint Univ. Cambridge DAMTP (Ph.D. thesis, 1991),
   and work in preparation
\bibitem[KaWa91]{KaWa91} {\sc H.G. Kausch, G.M.T. Watts},
%  {\em A Study of $\w$-Algebras using Jacobi Identities},
   Nucl. Phys. {\bf B354} (1991) 740
\bibitem[Nah91]{Nah91} {\sc W. Nahm},
%  {\em A proof of modular invariance},
   Int. J. Mod. Phys. {\bf A6} (1991) 2837, in Proc. Trieste July 1990 {\em
   Topological methods in quantum field theories}, World Scientific, 1991
\bibitem[Sal92]{Sal92} {\sc H. Saleur},
%  {\em Polymers and percolation in two dimensions and twisted $N=2$
%  supersymmetry},
   Nucl. Phys. {\bf B382} (1992) 486-531,
%  {\em Geometrical lattice models for $N=2$ supersymmetric theories in
%  two dimensions},
   Nucl. Phys. {\bf B382} (1992) 532-560
\bibitem[W$^2$H94]{WeWu94} {\sc Xiao-Gang Wen, Yong-Shi Wu,
   Yasuhiro Hatsugai},
%  {\em Chiral Operator Product Algebra and Edge Excitations of a Fractional
%  Quantum Hall Droplet},
   Nucl. Phys. {\bf B422}[FS] (1994) 476-494%,\\
%  {\sc Xiao-Gang Wen, Yong-Shi Wu},\\
%  {\em Chiral Operator Product Algebra Hidden in Certain Fractional Quantum
%  Hall Wave Functions},
%  Massachusetts Institute of Technology, Physics Department, Preprint 1994
  } %%% end footnotesize %%%
  \end{thebibliography}
  \end{document}